# Introduction to population dynamics and resource exploitation

Enrico Canuto and Daniele Mazza

Former Faculty, Politecnico di Torino

enrico.canuto@formerfaculty.polito.it

daniele.mazz@formerfaculty.polito.it

Version 1.1, September 2019

# Table of Contents

## Abstract

The paper was suggested by a brief note of the second author about the application of the Hubbert's curve to predict decay of resource exploitation. A further suggestion came from the interpretation of the Hubbert's curve in terms of a specific Lotka-Volterra (LV) equation. The link with population dynamics was obvious as logistic function and LV equation were proposed within the demography science field. Mathematical population dynamics has a history of about two centuries. The first principle and model of population dynamics can be regarded the exponential law of T. R. Malthus. In the XIX century, the Malthusian demographic model was first refined to include mortality rate by B. Gompertz. In the early XIX century the model was further refined by P-F. Verhulst by introducing the standard logistic function. The previous models only concern the population of a single species. In the early XX century, the American demographer A.J. Lotka and the Italian mathematician V. Volterra proposed a pair of state equations which describe the population dynamics of two competing species, the predator and the prey. This paper is concerned with the single and two-species fundamental equations: the logistic and LV equation. The paper starts with the generalized logistic equation whose free response is derived together with equilibrium points and stability properties. The parameter estimation of the logistic function is applied to the raw data of the US crude oil production. The paper proceeds with the Lotka-Volterra equation of the competition between two species, with the goal of applying it to resource exploitation. At the end, a limiting version of the LV equation is studied since it describes a competition model between the production rate of exploited resources and the relevant capital stock employed in the exploitation.

**Acronyms**

AS      asymptotically stable, asymptotic stability

DT      discrete-time

LV      Lotka-Volterra

LTI     linear time-invariant

RF      recovery factor

US      United States of America

# 1 Introduction

The paper was suggested by a brief note of the D. Mazza about the application of the Hubbert's curve (in other terms, the derivative of the logistic function) to predict decay of resource exploitation [1]. A further suggestion came from the interpretation of the Hubbert's curve in terms of a specific Lotka-Volterra (LV) equation by U. Bardi and A. Lavacchi [2]. The link with population dynamics was obvious as logistic function and LV equation were proposed within the demography science field.

Mathematical population dynamics has a history of about two centuries. The first principle and model of population dynamics can be regarded the exponential law [3] of T. R. Malthus [1766-1834], which is the free response of the state equation

$$\frac{dx(t)}{dt} = bx(t), \ x(t_0) = x_0 \ [\text{Vol}]. \tag{1}$$

In (1), $[\text{Vol}]^1$ denotes the unit of the population volume $x$, usually equal to the number of individuals or to their concentration in a region, and $b$ is the growth rate in $[\text{Vol/unit}]$, where $[\text{unit}]$ is a generic time unit. In the XIX century, the Malthusian demographic model was first refined [4] to include mortality rate (the decay rate of resource exploitation) by B. Gompertz [1779-1865] The corresponding state equation is

$$\frac{dx(t)}{dt} = bx(t)\ln\left(\frac{x_{max}}{x}\right), \ x(t_0) = x_0 \ [\text{Vol}], \tag{2}$$

where $x_{max} = \lim_{t \to \infty} x(t)$ is the asymptotic limit and $b$ is proportional to the initial growth rate. The free response, known as the Gompertz's function, is given by

$$x(t) = x_0 \exp\left(\ln(b/x_0)(1 - \exp(-bt))\right). \tag{3}$$

---

1 We will use the symbol [Vol] to denote the population volume measuring unit. An alternative symbol would be [Count].

By recalling the algebraic limit

$$\lim_{n \to \infty} \frac{1}{n}\left(1 - \left(\frac{x}{x_{max}}\right)^n\right) = \ln\left(\frac{x_{max}}{x}\right), \ n > 0, \tag{4}$$

replacement of $\ln(x_{max}/x)$ in (2) with the LHS of (4), provides the generalized logistic state equation ([5], also Richards' equation, from F.J. Richards who proposed the equation in 1959)

$$\frac{dx(t)}{dt} = \frac{b}{n}\left(1 - \left(\frac{x}{x_{max}}\right)^n\right)x(t), \ x(t_0) = x_0 \ [\text{Vol}]. \tag{5}$$

The free response of (5) for $n = 1$, $x(t_{max}) = x_{max}/2$ and $t > -\infty$, holds

$$x(t) = \frac{x_{max}}{1 + \exp(-b(t - t_{max}))} = \frac{x_{max}}{2}\left(1 + \tanh\frac{\exp(-b(t - t_{max}))}{2}\right). \tag{6}$$

The function in (6) was proposed in the early XIX century [6] by P-F. Verhulst [1804-1849] as an alternative refinement of (1) and is known as the *standard logistic function*, not to be confused with the symmetric logistic rate $\dot{x}(t)$. The term 'logistic' was created without explanation by P-F. Verhulst, presumably on the Ancient Greek λογῐστῐκή (the art of computing, a branch of the Ancient Greek Mathematics). For $t < t_{max}$ the population rate $\dot{x}(t)$ grows and for $t > t_{max}$ decays to zero. The symmetric logistic rate $\dot{x}(t)$ is also known as Hubbert's curve (and so it will be referred here) in the field of Earth's resource exploitation, since it was proposed by the American geologist M. King Hubbert in 1956 [7], thus establishing a link between population dynamics and resource exploitation.

Gompertz's and logistic equations (2) and (5) are autonomous state equations [17], since their solution only depends on initial conditions, and not on external perturbations and events. To obtain a finite asymptotic population limit $x_{max} = \lim_{t \to \infty} x(t)$, an intrinsic mortality term (known also as self-competition), i.e. the negative quadratic term in (5), is subtracted from+ the growth term, i.e. the positive and linear term in (5).

Gompertz's and Richards' equations only concern the population of a single species. In the early XX century, the American demographer A.J. Lotka (1880-1949) in 1920 [8] and the Italian mathematician V. Volterra (1860-1940) in 1926 [9] proposed a pair of state equations which describe the population dynamics of two competing species, the predator whose volume is denoted by $x_1$ and the prey whose volume is $x_2$:

$$\text{predator: } \dot{x}_1(t) = -\left(b_1 - a_{12}x_2(t)\right)x_1(t), \; x_1(0) = x_{10}, \\ \text{prey: } \dot{x}_2(t) = \left(b_2 - a_{21}x_1(t)\right)x_2(t), \; x_2(0) = x_{20} \tag{7}$$

The coefficients in (7) are positive. Each of the two equations follows the same template of the logistic equation (5) for $n = 1$, but the mortality term (the negative term in the second equation of (7)) of the prey and the growth term of the predator (the positive term in the first equation of (7)) are made extrinsic since they depend on the population volume of both species, thus accounting for their competition.

The LV equations (7) have been extended to include intrinsic mortality terms as in (5):

$$\text{predator: } \dot{x}_1(t) = -\left(b_1 - a_{12}x_2(t) + a_{11}x_1(t)\right)x_1(t), \; x_1(0) = x_{10}, \\ \text{prey: } \dot{x}_2(t) = \left(b_2 - a_{21}x_1(t) - a_{22}x_2(t)\right)x_2(t), \; x_2(0) = x_{20} \tag{8}$$

with positive coefficients. Equation (8) can be extended to a finite number $n$ of competing species. By denoting the volume vector with $\mathbf{x}$, and using the element-wise product symbol $\circ$, the $n-\dim$ LV equation becomes

$$\dot{\mathbf{x}}(t) = \left(\mathbf{b} + A\mathbf{x}(t)\right) \circ \mathbf{x}(t), \; \mathbf{x}(0) = \mathbf{x}_0, \tag{9}$$

where $\mathbf{b}$ is the vector of the intrinsic birth (positive) or death (negative) rates and the $n \times n$ matrix $A$ with components $a_{ij}$ accounts for relationships between species. The diagonal term $a_{ii}$ is usually negative and prevents the species to grow indefinitely. Two species $\{x_i, x_j\}$ are competing when $a_{ij} < 0$ and $a_{ji} < 0$. A species $x_i$ is a predator at the expense of $x_j$, when $a_{ij} > 0$ and $a_{ji} < 0$.

The two-dimensional equation (8) can be viewed as the linear case of the generic predator-prey equation [10]

$$\text{predator: } \dot{x}_1(t) = -b_1 x_1 + c_1(x_1, x_2)x_1, \; x_1(0) = x_{10}, \\ \text{prey: } \dot{x}_2(t) = x_2 b_2(x_2) - \eta_2 c_1(x_1, x_2)x_1, \; x_2(0) = x_{20} \tag{10}$$

where $b_1 > 0$ is the intrinsic predator mortality, $b_2(x_2) > 0$ is the per capita growth rate of the prey in absence of predator, $c_1(x_1, x_2) > 0$ is the per capita consumption rate of the predator, known as the trophic function (from the assumed Ancient Greek τρόφικος, relative to τροφή, food, nourishment). $c_1(x_1, x_2)$ determines, through the efficiency $\eta_2 > 0$, the mortality rate of the prey. A common assumption as in (7) is that the trophic function is only dependent on the prey volume, i.e. $c_1 = c_1(x_2)$. Based on experimental data, R. Arditi and L. R. Ginzburg suggested in [10] that actually $c_1$ depends on the ratio between prey and predator volumes, i.e. $c_1 = c_1 (x_2 / x_1)$, and that $c_1$ is a concave monotonic function tending to as asymptote.

Since the formalization in 1972 [11] by the British theoretical biologist J. Maynard Smith [1920-2004] (following a verbal statement of the American geneticist G.R. Price [1922-1975]) of the evolutionary stable strategies - a Nash equilibrium of the game theory which is evolutionary stable, in the sense that only natural selection protects the population evolution from small external perturbations - population dynamics has been complemented by the theory of games in order to model and explain population evolution within the evolutionary game theory [12].

This paper is just concerned with the single and two-species fundamental equations: the logistic and LV equation. Section 2 is devoted to the generalized logistic equation in (5). In Section 2.1, the free response (or logistic function) is derived together with equilibrium points and stability properties. The parameter estimation of the logistic function in Section 2.2 will be restricted to the standard logistic function and will be applied in Section 2.3 to the raw data of the US crude oil production from 1860 to 2018 [13].

Section 3 is devoted to the Lotka-Volterra (LV) equation of the competition between two species, with the goal of applying it to resource exploitation as suggested by [2]. To this end, the well-known LV equation (7) is studied in Section 3.1 pointing out the properties of the periodic free response and of the relevant closed trajectories around the so-called critical point. Section 3.2, which can be omitted at a first reading, derives, with the help of the literature [14], a method for computing the period of any LV closed trajectory. Section 3.3 provides a three-step method for estimating the six parameters of the LV equation (7), namely the four parameters $\{b_1, b_2, a_{12} / b_1, a_{21} / b_2\}$ of the equation and initial conditions. The first step aims to estimate the four parameters from explicit equations based on population measurements. At this stage, initial conditions are assumed to be equal to initial measurements. The second step refines the previous estimates by minimizing an error function. However, any two slightly different LV equations, although periodic and stable around their critical point, will drift from each other like two sine functions tuned on slightly different frequencies. To remove drift, the estimated equation must be tuned on the measurement data by means of a dynamic feedback, which is driven by the model error between measurements and the reconstructed response. This refinement is done by a third estimation step which is based on a stable discrete-time version of the LV equation, the topic being treated in Section 3.4. The section will point out that, although stable around the same critical equilibrium of the continuous-time equation, also the DT version will produce a drifting but bounded response. The drift can be removed by a dynamic feedback as already mentioned. Section 3.5 will apply the estimation procedure of Section 3.3 to the well-known 'lynx and hare data' collected around the early XX century in the northern Canadian forests [16], although recent experimental data suggest that the LV model is too simple. Finally, Section 3.6 studies the limiting version of the LV equation, which, as suggested in [2], describes a competition model between the production rate of exploited resources and the relevant capital stock employed in the exploitation. It is shown that the capital stock is the solution of a linear time-invariant (LTI) equation driven by the production rate. This allows to design a simple method for estimating equation parameters, namely, the capital obsolescence rate and the transformation coefficient between resources and capital. The method is applied to the sparse and uncertain data of the California gold rush, [18], [19], with some interesting result.

# 2 Single-species growth and mortality: the logistic equation

## 2.1 Logistic equation and free response

Let us consider the generalized logistic equation in (5) and apply, under $n > 0$, the variable change

$$z(t) = \left(x(t) / x_{\max}\right)^{-n} \geq 1,\tag{11}$$

and the relevant differential identity

$$dz = -n\left(\frac{x(t)}{x_{\max}}\right)^{-n-1}\frac{dx}{x_{\max}} \Rightarrow \frac{dx}{x_{\max}} = -\frac{x(t)}{x_{\max}}\frac{dz}{nz}.\tag{12}$$

Starting from (5) and by applying (11) and (12), we obtain, after some manipulations, a new equation as follows:

$$\frac{\dot{x}(t)}{x_{\max}} = \frac{b}{n}\left(1 - \left(\frac{x(t)}{x_{\max}}\right)^n\right)\frac{x(t)}{x_{\max}}, \ x(t_0) = x_0$$

$$-\frac{x(t)}{x_{\max}}\frac{\dot{z}(t)}{z\,n} = \frac{b}{n}\left(1 - z^{-1}\right)\frac{x(t)}{x_{\max}} \qquad .\tag{13}$$

$$\dot{z}(t) = -b\left(z(t) - 1\right), \ z(t_0) = z_0$$

*Free response.* The solution of the last equation in (13), which is LTI, is the following

$$z(t) = \exp\left(-b(t - t_0)\right)x_0 + b\int_{t_0}^t \exp\left(-b(t - \tau)\right)d\tau =$$

$$= 1 + (z_0 - 1)\eta(t - t_0)\tag{14}$$

where $\eta(t - t_0) = \exp\left(-b(t - t_0)\right)$ and the change of variable in (11) provides the free response of (5)

$$x(t) = \frac{x_{max}}{\left(1 + \left(\left(\frac{x_{max}}{x_0}\right)^n - 1\right)\eta(t - t_0)\right)^{1/n}}. \tag{15}$$

The response (15) reduces to the standard logistic function in (6) for $n = 1$ and $x_0 = x_{max}/2$.

*Equilibrium points and stability.* The last equation in (13) has a single natural equilibrium $\bar{z} = 1$, which is asymptotically stable (AS), since (14) provides $\lim_{t \to \infty} z(t) = 1$ and shows that the eigenvalue of the LTI perturbation equation is $-b$. Unlikely, the original equation (5) (the first equation in (13)) has two equilibrium points,

$$\bar{x}_0 = 0, \ \bar{x} = x_{max}. \tag{16}$$

The former equilibrium in (16), which corresponds to the divergent $\lim_{t \to \infty} z(t) = \infty$, is unstable, since the corresponding perturbation equation with $b/n > 0$ and $\delta x = x$ holds

$$\delta \dot{x}(t) = \frac{b}{n}\delta x(i), \ \delta x(0) = \delta x_0. \tag{17}$$

The second equilibrium $\bar{x} = x_{max}$ corresponds to $\bar{z} = 1$ and is left to the reader to prove that is AS under the LTI perturbation equation of (5) and to find the relevant eigenvalues.

*Generalized Hubbert's curve.* Last but not least, we derive from (15) a new expression of the logistic rate $\dot{x}(t)$ in (5), which holds:

$$\dot{x}(t) = b\frac{x_{max}\left(\left(\frac{x_{max}}{x_0}\right)^n - 1\right)\eta(t - t_0)}{\left(1 + \left(\left(\frac{x_{max}}{x_0}\right)^n - 1\right)\eta(t - t_0)\right)^{1/n+1}}. \tag{18}$$

The maximum value $\dot{x}_{max}$ and the argument $x^* = \arg\max_x \dot{x}(x)$ can be derived from the derivative identity $d\dot{x}(x)/dx = 0$, which from (5) holds:

$$\frac{d\dot{x}(x)}{dx} = \frac{b}{n}\left(1 - (n+1)\left(\frac{x}{x_{max}}\right)^n\right) = 0 \Rightarrow$$

$$\frac{x^*}{x_{max}} = \arg\max\left(\frac{\dot{x}(x)}{x_{max}}\right) = \left(\frac{1}{n+1}\right)^{1/n} . \tag{19}$$

$$\frac{\dot{x}_{max}}{x_{max}} = b\left(\frac{1}{n+1}\right)^{1/n+1}$$

The time argument $t_{max} = \arg\max_t \dot{x}(t)$ is obtained from the free response (15) and holds

$$t_{max} = t_0 \Rightarrow x^* = x_0 . \tag{20}$$

In the standard logistic case, corresponding to the Hubbert's curve, we obtain

$$\begin{aligned} x^* &= x_0 = x_{max} / 2 \\ \dot{x}_{max} &= b x_{max} / 4 \end{aligned} . \tag{21}$$

Replacement of $x_0 = x_{max} (n+1)^{-1/n}$ in (18), and of $t_0$ with $t_{max}$ provides the generalized and standard logistic rate expressions

$$\begin{aligned} \dot{x}(t) &= b \frac{x_{max} n \eta (t - t_{max})}{\left(1 + n \eta (t - t_{max})\right)^{1/n+1}} \\ \dot{x}(t) &= b \frac{x_{max} \eta (t - t_{max})}{\left(1 + \eta (t - t_{mqx})\right)^2} \ @ \ n = 1 \end{aligned} . \tag{22}$$

Figure 1 shows three profiles of the normalized logistic rate $\dot{x}(t) / x_{max}$ with $t_{max} = 10 \ unit$ and $b = 1 / unit$. The unitary exponent $n = 1$ separates two types of skew-symmetric profiles:

1) fast rising profiles with $n < 1$,
2) fast decaying profiles with $n > 1$.

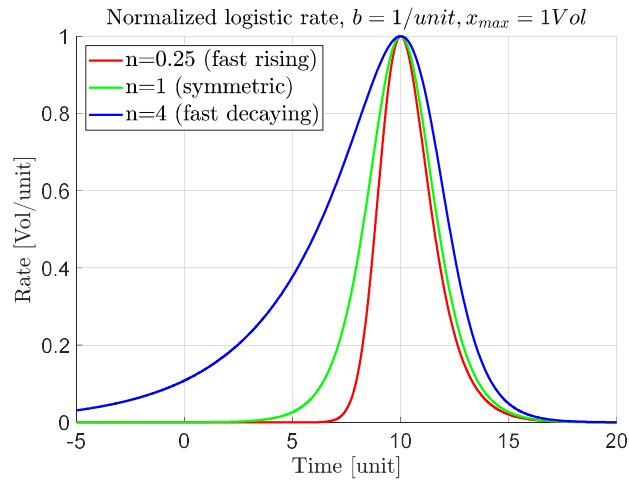

Figure 1.        Normalized logistic rate for different $n$.

## 2.2   Parameter estimation of the logistic rate

We will restrict to the standard logistic rate in (22) (the second equation), which is a function of three parameters collected in $\mathbf{p} = [x_{max}, b, t_{max}]$, where $x_{max}$ is the asymptotic population volume, $b$ is the growth rate coefficient and $t_0$ is the time instant of peak logistic rate. The functions in (6) and (22) suggest that knowledge of $t_{max}$ allows to estimate two parameters of $\mathbf{p}$, namely

$$x_{\max} = 2x(t_{\max})$$
$$b = 2\dot{x}(t_{\max}) / x(t_{\max}) = 4\dot{x}_{\max} / x_{\max} \qquad (23)$$

The critical step is the estimation of $t_{\max}$, i.e. of the time instant when $\dot{x}(t_{\max}) = \dot{x}_{\max}$ and the second time derivative $\ddot{x}(t)$ becomes zero. From (22) (second equation), the second derivative holds

$$\ddot{x}(t) = -x_{\max} b^2 \frac{\eta(t - t_{\max})(1 - \eta(t - t_{\max}))}{(1 + \eta(t - t_{\max}))^3} \Rightarrow \ddot{x}(t_{\max}) = 0 . \qquad (24)$$

When experimental data of the logistic rate $\dot{x}(t)$ are just increasing, $t_{\max}$ cannot be directly estimated from the raw data, but one can assume that either the last measurement corresponds to the maximum value of the second derivative or the relevant time instant can be obtained from the first derivative of the experimental data, after a convenient interpolation. Let us denote the this time instant with $t_{\inf}$ (it is the inflection point of the increasing logistic rate): it is the instant when the second derivative of the logistic rate, that is $\ddot{x}(t)$, becomes equal to zero. The third derivative holds

$$\dddot{x}(t) = -x_{\max} b^3 \eta(t - t_{\max}) \frac{\eta^2(t - t_{\max})^2 - 4\eta(t - t_{\max}) + 1}{(1 + \eta(t - t_{\max}))^4}, \qquad (25)$$

and by setting the derivative to zero, provides the interval $t_{\max} - t_{\inf} > 0$ as follows

$$\dddot{x}(t_1) = 0 \Rightarrow t_{\max} - t_{\inf} = \frac{1}{b}\left|\ln\left(\frac{1}{2 - \sqrt{3}}\right)\right| \cong \frac{1.32}{b} . \qquad (26)$$

From (26) we can obtain the value of the logistic rate:

$$\dot{x}(t_{\inf}) = bx_{\max} / 6 \qquad (27)$$

from the identity $\eta(t_{\inf} - t_{\max}) = 2 - \sqrt{3}$. Since identities (26) and (27) are just two equations with the three unknowns of $\mathbf{p} = [x_{\max}, b, t_{\max}]$, we need a third identity, possibly only involving experimental data of $\dot{x}(t)$. For instance, we can search the interval $|t_{\mathrm{half}} - t_{\max}|$ such that $\dot{x}(t_{\mathrm{half}}) = \dot{x}(t_{\inf}) / 2$ and $t_{\mathrm{half}} < t_{\inf}$. The relevant identity provides the second-degree equation

$$\eta_2^2 - 10\eta_2 + 1 = 0, \ \ \eta_2 = \exp(-b(t_{\mathrm{half}} - t_{\max})), \qquad (28)$$

whose solution, in the case of $t_{\mathrm{half}} < t_{\max}$, holds

$$t_{\max} - t_{\mathrm{half}} = \frac{1}{b}\left|\ln\left(\frac{1}{5 - \sqrt{24}}\right)\right| \cong \frac{2.29}{b} . \qquad (29)$$

By combining (26), (27) and (29), and by assuming to know $t_{\text{inf}} = \arg\max_t \ddot{x}(t)$ and $t_{\text{half}}$ from the identity $\dot{x}(t_{\text{half}}) = \dot{x}(t_{\text{inf}})/2$, one obtains a first-trial estimate $\mathbf{p}(0)$ of the parameters in $\mathbf{p}$, namely

$$\hat{b}(0) \cong \frac{2.29 - 1.32}{\widehat{t_{\text{inf}} - t_{\text{half}}}}$$
$$\hat{x}_{\max}(0) = 6\breve{y}(\hat{t}_{\text{inf}})/\hat{b}(0), \tag{30}$$
$$\hat{t}_{\max}(0) = \hat{t}_{\text{inf}} + \frac{1.32}{\hat{b}(0)}$$

where the concave mark $\cup$ denotes measurements, the convex mark $\cap$ estimates, and notations have been simplified by the identity $y(t) = \dot{x}(t)$. We assume also that the raw data $\breve{y}(t_i) = \breve{x}(t_i)$, $i = 0,...,N-1$, have been interpolated to obtain uniform sampling times $t_i = iT$. Given the $m$-th step parameter estimate $\hat{\mathbf{p}}(m)$, $m = 0,1,...$, the sampled measurements $\breve{y}(i) = \breve{x}(i)$ of the logistic rate and the estimated logistic rate $\hat{y}(i, \hat{\mathbf{p}}(m)) = \hat{x}(i, \hat{\mathbf{p}}(m))$, the estimation RMS error can be constructed as follows

$$\hat{s}(k) = \sqrt{\frac{1}{N} \sum_{i=0}^{N-1} \left(\breve{y}(i) - \hat{y}(i, \hat{\mathbf{p}}(m))\right)^2}, \tag{31}$$

and can be progressively minimized with respect to $\mathbf{p}$. The final RMS error is indicated by $\hat{s}$, and it can be normalized by the mean value $\underline{\breve{y}}(i)$ defined by

$$\underline{\breve{y}} = \frac{1}{N} \sum_{i=0}^{N-1} \breve{y}(i). \tag{32}$$

When the covariance of the measurement error is assumed to be known, the Cramer-Rao bound (the lower bound of the covariance matrix [17]) $P(\mathbf{p})$ of the parameter vector $\mathbf{p}$, can be estimated from the gradient $\mathbf{g}(t, \mathbf{p}) = \partial y(t, \mathbf{p})/\partial \mathbf{p}$ of the logistic rate in (22) (second row), computed at $\mathbf{p} = \hat{\mathbf{p}}$ (the parameter estimate) and at the sampling sequence $t_i = iT$, $i = 0,...,N-1$. The gradient holds

$$\mathbf{g}(t, \mathbf{p}) = \frac{\partial y(t, \mathbf{p})}{\partial \mathbf{p}} = y(t, \mathbf{p}) \left[ \frac{1}{x_{\max}} \quad \frac{1}{b}\left(1 - b(t - t_{\max})\frac{1 - \eta(t - t_0)}{1 + \eta(t - t_0)}\right) \quad b\frac{1 - \eta(t - t_0)}{1 + \eta(t - t_0)} \right]. \tag{33}$$

If we assume a statistical independent measurement error $\tilde{y}(i) = \breve{y}(i) - \hat{y}(i, \hat{\mathbf{p}})$ with known variance $\tilde{\sigma}_y^2(i)$, the estimated covariance is computed as

$$P(\hat{\mathbf{p}}) = \left( \sum_{i=0}^{N-1} \tilde{\sigma}_y^{-2}(i) \mathbf{g}(i, \hat{\mathbf{p}}) \mathbf{g}^T(i, \hat{\mathbf{p}}) \right)^{-1}, \tag{34}$$

where the variance $\sigma_{pk}^2$ of a single parameter $p_k$, $k = 1,2,3$ holds

$$\sigma_{pk}^2 = P_{kk}(\hat{\mathbf{p}}). \tag{35}$$

We will use also the normalized standard deviation (std) $\phi_{pk} = \sqrt{P_{kk}(\hat{\mathbf{p}})} / \hat{p}_k$. The covariance in (34) can be a-posteriori checked though a Monte Carlo run [17].

## 2.3 Application to US crude oil production

We apply the estimation method in Section 2.2 to a well-known set of raw data, the US annual production of crude oil from 1860 to 2019 [13], i.e. from the first successful use of a drilling rig in a well of Pennsylvania. The well was just drilled for producing oil by E.L. Drake and associates on August 27, 1959. The success of the Drake's well quickly led to oil drilling in other locations of United States and to a first wave of investments in oil drilling, refining and marketing. The principal product of oil was kerosene which replaced whale oil for illumination purposes. Before Drake's success, oil was an accidental byproduct of wells drilled for salt brine.

Generally speaking, petroleum reservoirs are permeable rock formations with minute porous spaces that contain trapped crude oil, natural gas or condensate (a low-density mixture of hydrocarbon liquids present in natural gas fields). The accumulated oil or gas is prevented to move upwards by an impermeable cap rock. Under favorable conditions, usually at the early stage of exploitation (the so-called primary recovery), oil production is made possible by drilling a well that produces a pressure difference between well and oil reservoir, thus allowing hydrocarbons to flow toward the well bore. The recovery factor (RF) expresses the amount of crude oil (or gas) which has been extracted as a percentage of the estimated hydrocarbon content of the reservoir. The primary recovery RF varies between 5 and 10%. After the primary recovery, pressure depletion alone becomes insufficient and injection of water or gas is required to keep the reservoir pressure at a convenient level for crude oil recovery. This process is known as the secondary recovery, which allows to raise the RF to a range between 15 and 40%. This implies that reservoirs are abandoned after the secondary recovery with, on the average, two thirds of their hydrocarbon content still left on the ground. The production raise in Figure 2 until the seventies of the XX century and the subsequent decay indicate the progressive exploitation and cessation of the secondary recovery in the US wells.

In the last twenty years, corresponding to the new increase of US crude oil production in Figure 2, a pair of unconventional exploitation processes have been implemented. The first one, referred to as tertiary recovery or end-of-reservoir process, aims to increase the RF of the abandoned wells to about the 60% of the estimated hydrocarbon content. The second process addresses rocks with low-permeability, like shale rocks (organic rich fine-grain sedimentary rocks), in which case hydraulic fracturing can stimulate the flow. It should be remarked that one of the first source of mineral oil, since prehistoric times, was oil shale, and oil shale industry, started in France in 1837 and steadily grew since before World War I, but declined, after World War II, due to accessibility of the Middle East crude oil.

Raw data and spline interpolation are shown in Figure 2. The interpolation time unit is 0.1 year. The ordinate is $10^6 \text{m}^3/\text{y}$, which was obtained from the conversion of the original thousand US barrels per day.

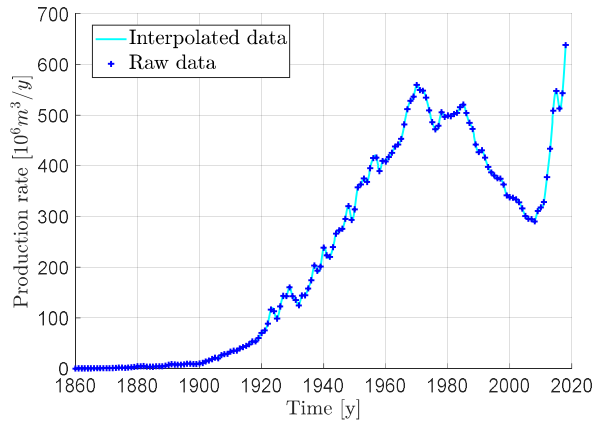

Figure 2. Interpolated and raw data of US crude oil production.

An immediate inspection of Figure 2 reveals that around the year 2000 a different kind of US crude oil production was implemented as explained above, which implies that at least two different logistic production rates $y_1(t) = \dot{x}_1(t)$ and $y_2(t) = \dot{x}_2(t)$ must be summed up, namely

$$y(t) = y_1(t) + y_2(t),  \tag{36}$$

where $y_1$ reached the maximum value around the year 1980, whereas $y_2$ is still increasing. A cascade estimation of the parameters of $\{y_1(t), y_2(t)\}$ has been done. The parameter vector $\mathbf{p}_1 = \left[ x_{\max,1}, b_1, t_{\max,1} \right]$ of $y_1$ has been estimated by restricting the raw data to $t_i \leq t_{\text{end},1} \cong 2005 \text{ y}$ (from inspection) and the parameter $\mathbf{p}_2 = \left[ x_{\max,2}, b_2, t_{\max,2} \right]$ of $y_2$ from the residuals of $\tilde{y}_1(i) = \bar{y}(i) - \hat{y}_1(i, \hat{\mathbf{p}}_1)$, but restricted to $t_i \geq t_{\text{start},2} \cong 2000 \text{ y}$. The initial parameter estimates $\hat{\mathbf{p}}_1(0)$ and $\hat{\mathbf{p}}_2(0)$ have been obtained with the methods of Section 2.2. More specifically $\hat{\mathbf{p}}_1(0)$ has been computed from (23) and the estimate of $t_{\max,1}$ from data inspection. Instead, $\hat{\mathbf{p}}_2(0)$ has been computed from (30). The final parameter estimates have been obtained by minimizing the estimation RMS error in (31). The MATLAB `fminsearch(.)` function has been employed.

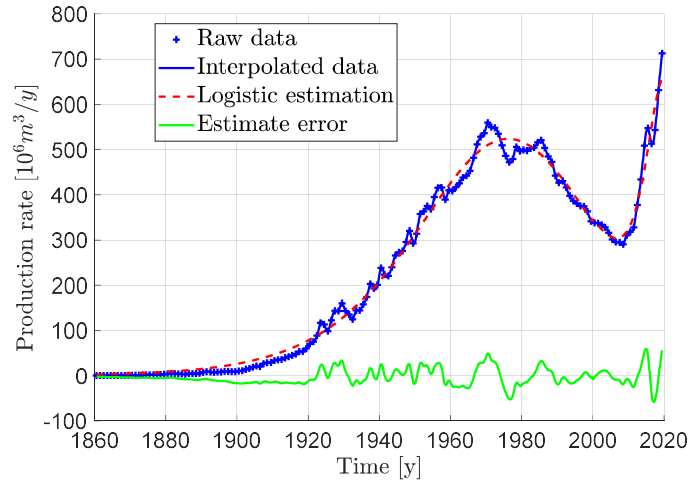

Figure 3.        Raw data, estimated logistic rate profile and estimation residuals.

Figure 3 compares interpolated raw data $\bar{y}(i)$ with the estimated profile $\hat{y}$ of the total function in (36). The estimation error profile $\tilde{y}(i)$ is also shown. Estimated parameters and their first-step estimated values are reported in Table 1.

| Table 1. US crude oil production estimation | | | | | |
|---|---|---|---|---|---|
| No | Parameter | Symbol | Unit | Value | Comments |
| First logistic rate $y_1$ | | | | | |
| 1 | Limit of the cumulative production | $\hat{x}_{max,1}(0)$ | $10^6\,m^3$ | 39800 | Initial estimate |
| 2 | Growth rate coefficient | $\hat{b}_1(0)$ | $y^{-1}$ | 0.0522 | Same |
| 3 | Rate peak time | $\hat{t}_{max,1}(0)$ | y | 1976.5 | Same |
| 4 | Limit of the cumulative production | $\hat{x}_{max,1}$ | $10^6\,m^3$ | 36600 | Final estimate |
| 5 | Growth rate coefficient | $\hat{b}_1$ | $y^{-1}$ | 0.0572 | Same |
| 6 | Rate peak time | $\hat{t}_{max,1}$ | y | 1976.2 | Same |
| 7 | Estimation RMS error | $\hat{s}_1$ | $10^6\,m^3/y$ | 16.6 | $t \leq t_{end,1}$ |
| 8 | Fractional RMS error | $\hat{s}_1 / \underline{\bar{y}}_1$ | fraction | 0.079 | same |

| 9 | Normalized parameter std | $\phi_{pk}$ | fraction | <0.005 | $k=1,2,3$ (each parameter) |
|---|---|---|---|---|---|
| Second logistic rate $y_2$ | | | | | |
| 10 | Limit of the cumulative production | $\hat{x}_{\max,2}(0)$ | $10^6\,\mathrm{m}^3$ | 7615 | Initial estimate |
| 11 | Growth rate coefficient | $\hat{b}_2(0)$ | $y^{-1}$ | 0.333 | Same |
| 12 | Rate peak time | $\hat{t}_{\max,2}(0)$ | y | 2022 | Same |
| 13 | Limit of the cumulative production | $\hat{x}_{\max,2}$ | $10^6\,\mathrm{m}^3$ | 7285 | Final estimate |
| 14 | Growth rate coefficient | $\hat{b}_2$ | $y^{-1}$ | 0.285 | Same |
| 15 | Rate peak time | $\hat{t}_{\max,2}$ | y | 2020.5 | Same |
| 16 | Estimation RMS error | $\hat{s}_2$ | $10^6\,\mathrm{m}^3/\mathrm{y}$ | 25.2 | $t_i \geq t_{\mathrm{start},2}$ |
| 17 | Normalized parameter std | $\phi_{pk}$ | fraction | <0.1 | $k=1,2,3$ (each parameter) |
| Total logistic profile | | | | | |
| 178 | Estimation RMS error | $\hat{s}$ | $10^6\,\mathrm{m}^3/\mathrm{y}$ | 17.7 | |
| 19 | Fractional RMS error | $\hat{s}/\bar{y}$ | fraction | 0.076 | |

Initial and final estimates of the first logistic profile $\bar{y}_1$ look very close since fairly all the profile is available from raw data. Something different occurs to the second profile $\bar{y}_2$, since the rate peak time in Table 1, row 14, has been anticipated, by the RMS error minimization, of two years, with respect to the first-step estimate in Table 1, row 11, obtained from (30) and the guess $t_{\mathrm{inf},2}=2018\,\mathrm{y}$. The difference is due to the intermediate peak at about the 2015 epoch, which likely hides an intermediate logistic profile. Only future production data may confirm or contradict the prediction.

The a priori fractional parameter standard deviation $\phi_{pk}$ has been computed from (35), by assuming a uniform measurement error std, which has been set equal to the RMS of the estimation error in Table 1, rows 7 and 16. The last assumption should be conservative since the estimation error, as one can see from Figure 3 (the green line), looks more a model error than a measurement error.

# 3 Two-species competition: the Lotka-Volterra equation

## 3.1 The Lotka-Volterra equation

### 3.1.1 Equation derivation and meaning

The well-known Lotka-Volterra (LV) equation of the predator-prey competition was independently proposed by the American demographer A.J. Lotka (1880-1949) in 1920 and by the Italian mathematician V. Volterra (1860-1940) in 1926. A.J. Lotka proposed the model to describe a hypothetical chemical reaction in which chemical concentrations oscillate [1]. V. Volterra proposed the model to explain the increase of predator fish (and the corresponding decrease of prey fish) in the Adriatic Sea after the World War I [9].

If we refer to predator-prey species competition, only two species are considered, predator $k = 1$ and prey $k = 2$. Their population volume or size [Vol] defines the state variables, which combine in the vector $\mathbf{x} = [x_1, x_2]$. Their per capita growth/decay rate is defined by

$$g_k(t) = \dot{x}_k(t) / x_k(t) = d \ln x_k / dt, \ k = 1, 2, \ x_k(t) \ge 0. \tag{37}$$

Let us assume a proportional decay (predator) and growth (prey) rate

$$\mathbf{g}(\mathbf{x}(t)) = [g_1(\mathbf{x}) \le 0, g_2(\mathbf{x}) \ge 0] = \mathbf{b} + A\mathbf{x}(t), \tag{38}$$

where $\mathbf{b} = [-b_1, b_2]$, $b_k \ge 0$ [Vol/s] is the intrinsic growth/decay vector, decay for predator and growth for the prey, i.e. the growth/decay rate in absence of competition. The matrix $A$ [(Vol/s)/Vol], the interaction matrix, describes the effect of the population volume on the per capita growth. The resulting autonomous equation

$$\dot{\mathbf{x}}(t) = \mathbf{g}(\mathbf{x}(t)) \circ \mathbf{x}(t), \ \mathbf{x}(0) = \mathbf{x}_0, \tag{39}$$

where $\circ$ denotes the Hadamard element-wise product, is known as the Kolmogorov's predator-prey equation. In the LV equation, only mutual interactions are accounted for, which leads to

$$A = \begin{bmatrix} 0 & a_{12} \\ -a_{21} & 0 \end{bmatrix}, \ a_{ij} \ge 0. \tag{40}$$

Let us remark that $A$ in (40) has the same form of the state matrix of an undamped oscillator. By combining (37), (38) and (40), the LV state equation reads as

$$\dot{\mathbf{x}}(t) = (\mathbf{b} + A\mathbf{x}(t)) \circ \mathbf{x}(t), \ \mathbf{x}(0) = \mathbf{x}_0, \ \mathbf{x}(t) \ge 0. \tag{41}$$

The equation components are as follows:

$$\text{predator: } \dot{x}_1(t) = -\big(b_1 - a_{12}x_2(t)\big)x_1(t), \ x_1(0) = x_{10}, \ x_1(t) \geq 0$$
$$\text{prey: } \dot{x}_2(t) = \big(b_2 - a_{21}x_1(t)\big)x_2(t), \ x_2(0) = x_{20}, \ x_2(t) \geq 0 \quad , \tag{42}$$

where all the coefficients and variables are nonnegative.

The equation block-diagram is shown in Figure 4. The three loops, the decay and growth loops around the predator and prey state variables, respectively, and the overall oscillating loop from predator to prey are shown. The dashed lines may account for environmental actions.

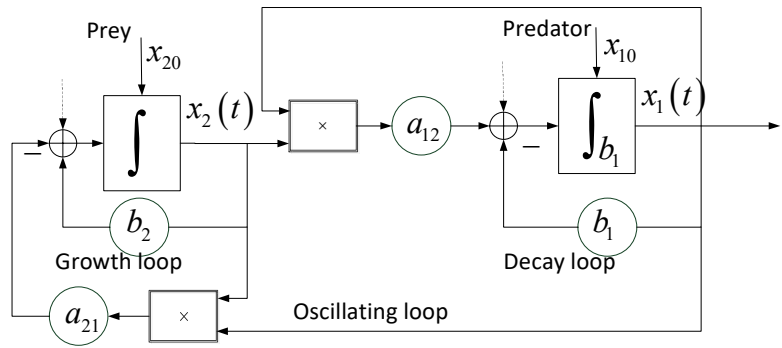

Figure 4.        Block-diagram of the LV equation (42).

### 3.1.2   Equilibrium points and conservation equation

Equation (42) has a pair of equilibrium points: the zero equilibrium $\overline{\mathbf{x}}_0 = [0,0]$ and the *critical (equilibrium) point*

$$\overline{\mathbf{x}} = [\overline{x}_1, \overline{x}_2] = [b_2/a_{21}, b_1/a_{12}]. \tag{43}$$

To proceed to the analysis, (42) is rewritten in terms of the normalized population volumes

$$\mathbf{z} = [z_1, z_2] = [x_1/\overline{x}_1, x_2/\overline{x}_2], \tag{44}$$

which transformation provide the following nonlinear equation

$$\dot{\mathbf{z}}(t) = \begin{bmatrix} z_1(t) & 0 \\ 0 & z_2(t) \end{bmatrix} \begin{bmatrix} 0 & b_1 \\ -b_2 & 0 \end{bmatrix} \big(\mathbf{z}(t) - \overline{\mathbf{z}}\big), \ \mathbf{z}(0) = \mathbf{z}_0 \quad .$$
$$\mathbf{x}(t) = [z_1(t)\overline{x}_1, z_2(t)\overline{x}_2], \ \overline{\mathbf{z}} = [1,1] \tag{45}$$

The LV equation (45) cannot be integrated in closed-form, but a constant integral is found by multiplying the first equation in (45) by $b_2(z_1-1)/z_1$, the second one by $b_1(z_2-1)/z_2$ and by summing their products to get the following identity

$$b_2(z_1-1)\,\dot{z}_1/z_1 + b_1(z_2-1)\dot{z}_2/z_2 = 0, \tag{46}$$

which is equivalent to the differential equation

$$\frac{d}{dt}\big(b_2\left(z_1 - \ln z_1\right) + b_1\left(z_2 - \ln z_2\right)\big) = 0 . \tag{47}$$

In turn, the differential equation (47) can be integrated to provide the constant function

$$\begin{aligned}V(\mathbf{z}) &= b_2\big(z_1(t) - \ln z_1(t)\big) + b_1\big(z_2(t) - \ln z_2(t)\big) = V(\mathbf{z}_0) = \\ &= b_2\left(z_{10} - \ln z_{10}\right) + b_1\left(z_{20} - \ln z_{20}\right) > 0,\ t \ge 0\end{aligned} , \tag{48}$$

which is clearly positive. Equation (48) is a *conservation equation*, which is typical of oscillatory systems: at any time, the integral $V(\mathbf{z}(t)) = V(\mathbf{z}_0)$ is conserved. This implies that the trajectories of (45), lying in the positive quadrant $\vartheta_+ = \{\mathbf{z}(t) \ge 0\}$ of the state space, are periodic. Prey population increases when predator population decreases and vice versa.

### 3.1.3 Other proofs that trajectories are closed

The gradient and the Hessian matrix of $V(\mathbf{z})$

$$\begin{aligned}\frac{\partial V(\mathbf{z})}{\partial \mathbf{z}} &= \Big[\, b_2\left(1 - 1/z_1\right) \quad b_1\left(1 - 1/z_2\right)\,\Big] \\ \frac{\partial^2 V(\mathbf{z})}{\partial \mathbf{z}^2} &= \mathrm{diag}\big(b_2/z_1^2, b_1/z_2^2\big) > 0\end{aligned} , \tag{49}$$

show that $V(\overline{\mathbf{z}} = [1,1])$ is the unique minimum value of $V(\mathbf{z})$ in the positive quadrant, and all the trajectories are contour curves of $V(\mathbf{z})$ around $V(\overline{\mathbf{z}}) = b_2 + b_1 = \overline{E}$. This implies that $U(\mathbf{z}) = V(\mathbf{z}) - V(\overline{\mathbf{z}}) \ge 0$, i.e. that

$$\begin{aligned}U(\mathbf{z}) &= b_2\big(z_1(t) - \ln z_1(t) - 1\big) + b_1\big(z_2(t) - \ln z_2(t) - 1\big) \\ U(\overline{\mathbf{z}}) &= 0,\ E = U(\mathbf{z}_0) \ge 0\end{aligned} , \tag{50}$$

is *a candidate Lyapunov function* and $E$ can be defined as the 'orbit energy' in [Vol/s] units. The energy has been set to zero in the critical point. The time derivative has been already computed in (46) and holds zero

$$\dot{U}(\mathbf{z}(t)) = b_2\left(z_1 - 1\right)\dot{z}_1/z_1 + b_1\left(z_2 - 1\right)\dot{z}_2/z_2 = 0 , \tag{51}$$

which confirms that trajectories are closed around the critical point $\overline{\mathbf{x}} \ne 0$.

A second proof comes from the use of *Poincaré map* (see Section 5.1). To this end, let us rewrite $V(\mathbf{z}) = V(\mathbf{z}_0)$ with the help of (48), as follows

$$b_2 \ln\left(z_1/z_{10}\right) + b_1 \ln\left(z_2/z_{20}\right) = b_2\left(z_1 - z_{10}\right) + b_1\left(z_2 - z_{20}\right) , \tag{52}$$

and let us select the line $z_k = z_{k0},\ k = 1, 2$, as the return surface $\Sigma_k$. With the notation $\xi_k = z_k/z_{k0}$, the identity (52) simplifies to

$$\ln \xi_{j \neq k} = z_{j0}\left(\xi_j - 1\right), \tag{53}$$

and shows that the solution $\xi_j = 1 \Rightarrow z_j = z_{j0}$ always exists and indicates that the return point coincides with the initial point as for closed trajectories, in agreement with the Poincaré map $P(\mathbf{z}_0) = \mathbf{z}_0$. The unit value $\xi_j = 1$ implies that the crossing point is unique and is just the initial point. In other terms, the trajectory is tangent to the line $\Sigma_k$. Intermediate crossing points exist and depend on the value of $\xi_{j0} \neq 1$. Under $\xi_{j0} > 1$, the intermediate crossing point holds $\{z_{k0}, \hat{z}_{j \neq k} < z_{j0}\}$, whereas under $\xi_{j0} < 1$ it holds $\{z_{k0}, \hat{z}_{j \neq k} > z_{j0}\}$. A significant property is that the intermediate crossing point component $\hat{z}_{j \neq k}$ does not depend on $z_{k0}$ and thus on $\Sigma_k$.

A third proof comes by converting $U(\mathbf{z})$ in (50) into a *Hamiltonian function*. The nonlinear state transformation $s_k = \ln z_k \Leftrightarrow \exp(s_k) = z_k$, $k = 1,2$ changes $U(\mathbf{z})$ into a Hamiltonian function as follows

$$H(\mathbf{s}) = b_2\left(\exp s_1 - s_1(t) - 1\right) + b_1\left(\exp s_2(t) - s_2(t) - 1\right) \geq 0. \tag{54}$$

To prove that $H(\mathbf{s})$ is Hamiltonian, let us transform the state equations (45) in terms of $\mathbf{s} = [s_1, s_2]$:

$$\begin{aligned}
\dot{z}_1(t) &= b_1 z_1(z_2 - 1) \Rightarrow \dot{s}_1(t) = b_1\left(\exp(s_2) - 1\right) \Rightarrow \\
\dot{z}_2(t) &= -b_2 z_2(z_1 - 1) \Rightarrow \dot{s}_2(t) = -b_2\left(\exp(s_1) - 1\right) \Rightarrow \\
\dot{s}_1(t) &= \frac{\partial H(\mathbf{s})}{\partial s_2}, \ s_1(0) = s_{10} \\
\dot{s}_2(t) &= -\frac{\partial H(\mathbf{s})}{\partial s_1}, \ s_2(0) = s_{20}
\end{aligned}, \tag{55}$$

where the last two state equations constitute a Hamiltonian system. Since the Hamiltonian in (54) does not explicitly depend on time, the energy conservation law states that the Hamiltonian is equal to the 'constant energy' of the orbit, namely $E = H(\mathbf{s}_0)$.

### 3.1.4   Average population

A further meaning of the critical point $\overline{\mathbf{x}} = [\overline{x}_1, \overline{x}_2] \neq 0$ is that of providing the *average population volume* along any closed trajectory. In fact, let us consider a generic trajectory of period $T$ such that $\mathbf{z}(T) = \mathbf{z}_0$ and from (45) the equation

$$\frac{d}{dt}\ln \mathbf{z}(t) = \begin{bmatrix} \dot{z}_1 / z_1 \\ \dot{z}_2 / z_2 \end{bmatrix} = \begin{bmatrix} 0 & b_1 \\ -b_2 & 0 \end{bmatrix}(\mathbf{z}(t) - \overline{\mathbf{z}}), \ \mathbf{z}(0) = \mathbf{z}_0. \tag{56}$$

Integration of (56) along the interval $[0, T]$ provides

$$\ln \mathbf{z}(T) - \ln \mathbf{z}_0 \begin{bmatrix} 0 & b_1 \\ -b_2 & 0 \end{bmatrix} \left( \int_0^T \mathbf{z}(\tau)\, d\tau - \overline{\mathbf{z}}T \right) = 0, \tag{57}$$

and consequently, as expected,

$$\overline{\mathbf{z}} = \frac{1}{T} \int_0^T \mathbf{z}(\tau)\, d\tau \Leftrightarrow \overline{\mathbf{x}} = \frac{1}{T} \int_0^T \mathbf{x}(\tau)\, d\tau. \tag{58}$$

### 3.1.5 Perturbation equation

The perturbation equation about the critical point is expected to be oscillatory. Let us define the perturbation $\delta \mathbf{z} = \begin{bmatrix} \delta z_1 = z_1 - 1, \delta z_2 = z_2 - 1 \end{bmatrix}$ of the normalized state $\mathbf{z}$ in (44). Equation (45) can be rewritten as

$$\delta \dot{\mathbf{z}}(t) = \begin{bmatrix} 1 + \delta z_1(t) & 0 \\ 0 & 1 + \delta z_2(t) \end{bmatrix} \begin{bmatrix} 0 & b_1 \\ -b_2 & 0 \end{bmatrix} \delta \mathbf{z}(t), \ \delta \mathbf{z}(0) = \delta \mathbf{z}_0. \tag{59}$$

The LTI part of the state matrix in (59) has a pair of imaginary eigenvalues $\lambda_{1,2} = \pm j \sqrt{b_1 b_2} = \pm j \omega_{LV} = \pm j 2\pi f_{LV}$, which do not depend on the prey-predator interaction coefficients $\{a_{12}, a_{21}\}$. The relevant period $P_{LV} = 1/f_{LV}$ is just the period of the small orbits close to the critical point, since the orbit period will depend on the initial state $\mathbf{z}_0$ encoded in the orbit energy $E \geq 0$ defined in (50).

### 3.1.6 Graphical plots

Figure 5 shows the time profile of the pair (predator, prey) by using the $\mathbf{x} = \begin{bmatrix} x_1, x_2 \end{bmatrix}$ variables of equation (7) and starting from the following initial condition:

$$\mathbf{x}_0 = \begin{bmatrix} \overline{x}_1 (1 + 3/4), \overline{x}_2 (1 + 3/2) \end{bmatrix}. \tag{60}$$

The equation coefficients and the population average have values:

$$\mathbf{b} = \begin{bmatrix} -2, 4 \end{bmatrix}, \ A = \begin{bmatrix} 0 & 0.25 \\ -1 & 0 \end{bmatrix}, \ \overline{\mathbf{x}} = \begin{bmatrix} 4, 8 \end{bmatrix}. \tag{61}$$

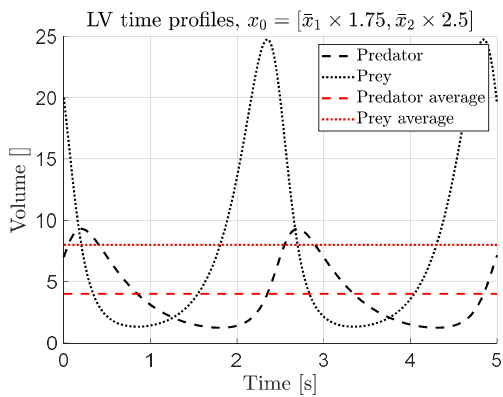
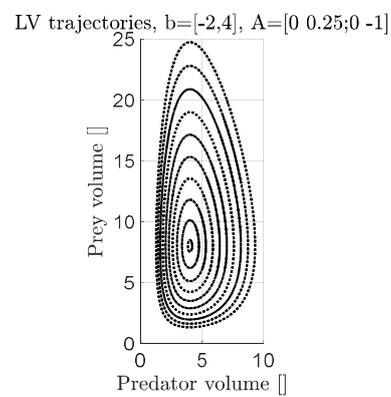

Figure 5.    Left: time profiles of the largest trajectory on the right figure. Right: LV trajectories about the critical point (equal to the population average) [4,8].

### 3.1.7    The data that suggested V. Volterra equation

The weakness of the LV equation is of being *autonomous*, thus incapable of accounting for external perturbations. In fact, the LV equation suggested by V. Volterra [9], could not per se explain, as the output of a cause-effect phenomenon, the predator fish fraction (mainly of the clade Selachimorpha, like sharks) increase in Northern Adriatic Sea during and after the World War I, since the cause should have been related to war and specifically to a diminished fishing activity. The record of the predator fish fraction over the whole caught fish per year [16] is shown in Figure 6.

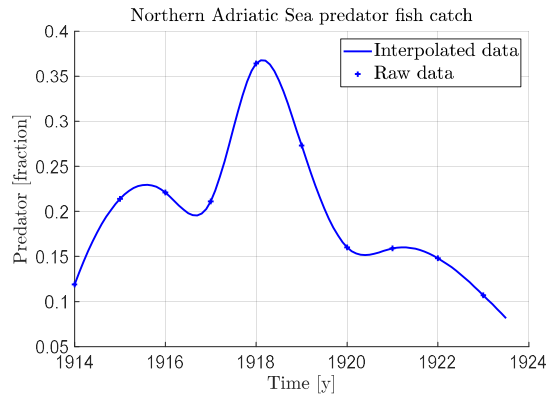

Figure 6.    Raw and interpolated data of the predator fish fraction.\

The available data are scarce and moreover only the predator fish fraction is available, which implies that the LV equation (7) should be rewritten as follows. Let $X = x_1 + x_2$ the total amount of caught fish per year, and $\xi_k = x_k / X$, $k = 1,2$ the fraction of predator and prey. It is straightforward to rewrite (7) as follows

$$\text{predator: } \dot{\xi}_1(t) = -\left(b_1 + \beta(t)\right)\xi_1(t) + a_{12}X(t)\xi_1(t)\left(1 - \xi_1(t)\right), \ \xi_1(0) = \xi_{10},$$
$$\text{total: } \dot{X}(t) = \beta(t)X(t), \ X(0) = X_0 \tag{62}$$

where the intrinsic rate perturbation $\beta(t)$ and the initial fish population $X_0$ are not available. A similar equation holds for the prey

$$\text{prey: } \dot{\xi}_2(t) = \left(b_2 - \beta(t)\right)\xi_2(t) + a_{21}X(t)\xi_2(t)\left(1 - \xi_2(t)\right), \ \xi_2(0) = \xi_{20}, \tag{63}$$

where, now, the rate perturbation $\beta(t)$ due to time varying fishing, affects $b_1$ with the opposite sign of $b_2$ in (62). In conclusion, fishing reduction implies $\beta(t) < 0$, and consequently an increase of the prey intrinsic growth rate, now $b_2 - \beta(t)$, and a reduction of the predator intrinsic mortality rate, now $b_1 + \beta(t)$. Scarce and incomplete data prevent further considerations and numerical estimation.

## 3.2 Orbit period

### 3.2.1 Introduction

This section may be omitted at a first reading. We have already noticed that the perturbation period $P_{LV} = 2\pi / \sqrt{b_1 b_2}$ is just the period of the small orbits about the critical point $\overline{\mathbf{x}} = [b_2 / a_{21}, b_1 / a_{12}]$. The issue is to find the expression of a generic period $P(E)$ as a function of the initial conditions, which are encoded in the orbit energy $E$. Some equivalent expressions can be found in the literature as reported by S-D. Shih in [14], who in addition derived the same expression as that of F. Rothe [15], but in a simpler way.

### 3.2.2 S-D. Shih derivation

We report the steps of the S-D. Shih's derivation. Let us begin from $U(\mathbf{z})$ in (50), which is set equal to the 'orbit energy' $U(\mathbf{z}_0) = E \geq 0$ and then exponentiated as follows

$$
\begin{aligned}
&\exp\left(\left(z_1 - \ln z_1 - 1\right) + \left(b_1 / b_2\right)\left(z_2 - \ln z_2 - 1\right)\right) = \exp\left(E / b_2\right) \Rightarrow \\
&\Rightarrow \frac{\left(\exp(z_2) / z_2\right)^{b_1 / b_2}}{z_1 \exp(-z_1)} = \exp\left(E / b_2 + 1 + b_1 / b_2\right)
\end{aligned}
\tag{64}
$$

As a second step, let us solve for $z_1$ by separating the variables as follows

$$
\begin{aligned}
&h(z_1) = \left(\exp z_2 / z_2\right)^{b_1 / b_2} / \exp\left(\left(\overline{E} + E\right) / b_2\right) = \eta(z_2, E), \\
&h(z_1) = z_1 \exp(-z_1), \ z_1 \geq 0
\end{aligned}
\tag{65}
$$

where $h(z)$ is shown in Figure 7, $z \geq 0$ being a generic nonnegative real, and $\overline{E}$ is the energy at the critical point. Since $h(z)$ is not one-to-one invertible, it must be written as the sum of two invertible functions $h(z) = h_0(z) + h_1(z)$, which are defined as follows

$$
\begin{aligned}
&h_0(z) = z \exp(-z), \ 0 \leq z < 1 \\
&h_1(z) = z \exp(-z), \ 1 \leq z
\end{aligned}
\tag{66}
$$

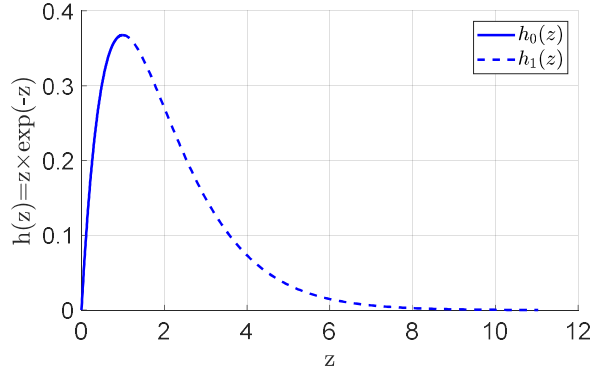

Figure 7.        The components of $h(z)$.

The range of $h(z)$, $z \geq 0$ is bounded and holds $0 \leq h(z) \leq e^{-1} < 1$. With the help of the notation $g_k(u) = h_k^{-1}(u)$, we can formally write

$$z_1 = g_0(\eta(z_2, E)), \ 0 \leq z_1 < 1$$
$$z_1 = g_1(\eta(z_2, E)), \ 1 \leq z_1 \qquad , \qquad (67)$$

where $\eta(z_2, E)$ has been defined in (65). Equations in (67) need to be completed with their range $0 \leq z_{2,\min} \leq z_2 \leq z_{2,\max}$, which is obtained from the inequalities

$$0 \leq \eta(z_2, E) = (\exp z_2 / z_2)^{b_1/b_2} / \exp((\bar{E} + E)/b_2) \leq \exp(-1) \Rightarrow$$
$$\Rightarrow 0 < \exp\left(-1 - \frac{E}{b_1}\right) \leq z_2 \exp(-z_2) \leq \exp(-1) \qquad . \qquad (68)$$

The second row inequalities have two solutions $\{z_{2,\min} \leq 1, z_{2,\max} \geq 1\}$, which coalesce into $\{1,1\}$ for $z_2 = \bar{z}_2 = 1$, and are defined by

$$z_{2,\min}(E) = g_0(\exp(-1 - E / b_1))$$
$$z_{2,\max}(E) = g_1(\exp(-1 - E / b_1)) \qquad . \qquad (69)$$

By replacing $\exp(z_2) / z_2 = \exp(1 + E / b_1)$ into (64) one finds that the pair $\{z_{2,\min}, z_{2,\max}\}$ corresponds to the orbit points $\{P_0, P_2\}$ with coordinates $\mathbf{z}_0 = [1, z_{2,\min}]$ and $\mathbf{z}_2 = [1, z_{2,\max}]$, as Figure 8 shows. The same holds for the pair $\{z_{1,\min}(E), z_{1,\max}(E)\}$ and the points $\{P_1, P_3\}$ with coordinates $\mathbf{z}_1 = [z_{1,\min}, 1]$ and , $\mathbf{z}_3 = [z_{1,\max}, 1]$. This implies that the period computation can be written as the sum of time-integrals along the four orbit branches $\{P_k, P_{\mathrm{mod}(k+1,4)}\}, k = 0,1,2,3$, once the relation between the time differential $dt$ and the state differential either $dz_2$ (in this case) or $dz_1$ has been found.

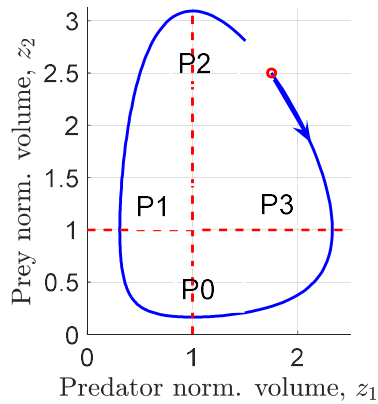

Figure 8.    The clockwise normalized orbit starting from the red point.

The time differentials are obtained from (45) by replacing either $z_1$ with $z_{1k} = g_k(z_2)$, $k = 0,1$ in the second equation or $z_2$ with $z_{2k} = g_k(z_1)$, $k = 0,1$ in the first equation, which provides

$$dt_{k,1} = \frac{dz_1}{z_1 b_1 \left( g_k \left( \eta(z_1, E) \right) - 1 \right)}$$
$$dt_{k,2} = \frac{dz_2}{z_2 b_2 \left( 1 - g_k \left( \eta(z_2, E) \right) \right)} \tag{70}$$

Using the second differential, the orbit period $P(E)$ in [s] units, is the solution of the following integral

$$P(E) = \int_{z_{2,\min}(E)}^{1} \left( \frac{1}{1 - g_0(\eta(z_2, E))} - \frac{1}{1 - g_1(\eta(z_2, E))} \right) \frac{dz_2}{b_2 z_2} +$$
$$+ \int_{1}^{z_{2,\max}(E)} \left( \frac{1}{1 - g_0(\eta(z_2, E))} - \frac{1}{1 - g_1(\eta(z_2, E))} \right) \frac{dz_2}{b_2 z_2} \tag{71}$$

where $g_0(\cdot)$ applies to $z_1 < 1$, $g_1(\cdot)$ to $z_1 \geq 1$ and the minus sign to $dz_2 < 0$. Following [14], the above integral can be converted into a convolution integral through the substitution

$$z_2(\eta) \exp(-z_2(\eta)) = \exp(-1 - \eta / b_1), \tag{72}$$

which entrains the following transformations

$$z_2 = g_k \left( \exp(-1 - \eta / b_1) \right), \ g_0 \Leftrightarrow z_2 < 1, \ g_1 \Leftrightarrow z_2 \geq 1$$
$$\eta(z_2, E) = \exp\left( -1 - \frac{E - \eta}{b_2} \right) \tag{73}$$

and the new integration limits and differential

$$z_2 = z_{2,\min} = z_{2,\max} \Rightarrow \eta = E$$
$$z_2 = 1 \Rightarrow \eta = 0 \tag{74}$$
$$dz_2 = -\frac{z_2 d\eta}{b_1(1 - z_2)}$$

Replacement of (73) and (74) in (71) and some manipulations allow us to rewrite (71) as the following convolution integral

$$P(E) = \frac{1}{\omega_{LV}^2} \int_0^E G\left(\frac{\eta}{b_1}\right) G\left(\frac{E - \eta}{b_2}\right) d\eta \left[\frac{\text{Vol/s}}{\text{Vol/s}^2} = \text{s}\right], \tag{75}$$

where the dimensionless function to be integrated holds

$$G(\sigma) = \frac{1}{1 - g_0(\exp(-1 - \sigma))} - \frac{1}{1 - g_1(\exp(-1 - \sigma))}. \tag{76}$$

In the numerical integration of (75) written as

$$P(E) \cong \frac{E_{\max}}{\omega_{LV}^2} \sum_{i=0}^{N-1} G\left(\frac{\eta(i)}{b_1}\right) G\left(\frac{E - \eta(i)}{b_2}\right),$$
$$\eta(i) = \left(i + \frac{1}{2}\right) \frac{E_{\max}}{N} \tag{77}$$

one should pay attention that $G(\sigma)$ is singular for $\sigma \to 0$, the first-order approximation being

$$\lim_{\sigma \to 0} G(\sigma) = 2/\sqrt{\sigma}. \tag{78}$$

This implies that, unlike in (77), the integration step should become smaller for $\eta \to 0$ and $\eta \to E$. Figure 9 shows a typical argument of the summation in (77) with a fractional energy quantization $\Delta\eta / E \cong 0.001$. The fractional error between the period computed by (77) and the period estimated by simulation is about $0.006$. A fractional error larger than the quantization is likely due to the uniform quantization of the numerical integration.

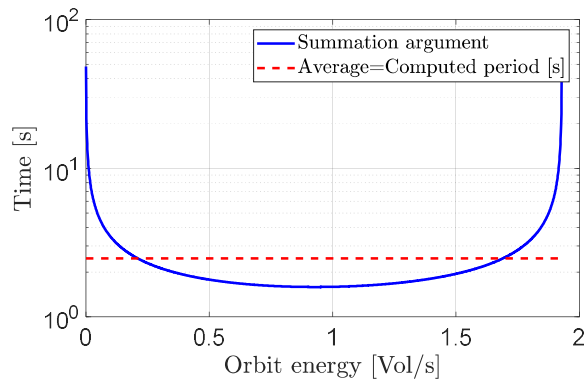

Figure 9.    Typical argument of (77) and the average (the computed period).

## 3.3 Estimation of the LV equation parameters

### 3.3.1 Method

Estimation of LV parameters from experimental data is organized into three steps.

1) Fist we perform the estimation of the population averages $\{\bar{x}_1, \bar{x}_2\}$ and of the rate coefficients $\{b_1, b_2\}$ from experimental data $\{\bar{x}_1(i), \bar{x}_2(i)\}$, $i = 0, ..., N-1$, where uniform sampling, i.e. $t_i = iT$, is assumed. If necessary, the samples can be interpolated through splines, especially in view of the population rate computation $\{\breve{x}_1(i), \breve{x}_2(i)\}$ and of the uniform sampling. Let us collect the previous estimates and the initial conditions in the vector

$$\hat{\mathbf{p}}(0) = \left[ \bar{x}_{10}, \hat{b}_1, \hat{a}_{12} = \hat{b}_1 / \widehat{\bar{x}}_2, \bar{x}_{20}, \hat{b}_2, \hat{a}_{21} = \hat{b}_2 / \widehat{\bar{x}}_1 \right]. \tag{79}$$

2) The second step performs a refinement of the previous estimate through the progressive minimization of an error functional $\eta_m\left(\{\bar{x}_1(i), \bar{x}_2(i)\}, \{x_{1m}(i), x_{2m}(i)\}; \hat{\mathbf{p}}(m)\right)$, $m = 0, 1, ...,$ between raw (interpolated) data $\{\bar{x}_1(i), \bar{x}_2(i)\}$ and simulated samples $\{x_{1m}(i), x_{2m}(i)\}$, where the subscript $m \geq 0$ denotes a generic optimization step, until some convergence criterion is met. The simulated samples derive from the integration of (42) based on the estimated parameter vector $\hat{\mathbf{p}}(m)$, $m \geq 0$. The error functional is defined as the RMS of the normalized error between raw and simulated samples as follows.

$$\eta_m(\cdot) = \sqrt{\frac{1}{N} \sum_{i=0}^{N-1} \sum_{k=1}^{2} \left( \frac{\bar{x}_k(i) - x_{km}(i; \hat{\mathbf{p}}(m))}{\widehat{\bar{x}}_k} \right)^2}. \tag{80}$$

The normalization is fixed and is done with respect to the raw data average $\{\widehat{\bar{x}}_1, \widehat{\bar{x}}_2\}$ obtained at the first step. The MATLAB `fminsearch(.)` function will be employed, as already done in Section 2.3.

3) The last step interprets the final residual errors $\tilde{x}_k(i) = \bar{x}_k(i) - x_k(i; \hat{\mathbf{p}})$, $k = 1, 2$ of the second step as the result of an external inputs $u_k(i)$ to the autonomous LV equation (42), which therefore is rewritten as

$$\begin{aligned} \text{predator: } &\dot{x}_1(t) = -\left(b_1 - a_{12}x_2(t)\right)x_1(t) + u_1(t), \ x_1(0) = x_{10}, \ x_1(t) \geq 0 \\ \text{prey: } &\dot{x}_2(t) = \left(b_2 - a_{21}x_1(t)\right)x_2(t) + u_2(t), \ x_2(0) = x_{20}, \ x_2(t) \geq 0 \end{aligned}. \tag{81}$$

To this end, a DT state equation of (81) is necessary as it will be done in Section 3.4. secondary, $u_k(i)$ must be estimated as a dynamic feedback driven by $\tilde{x}_k(i)$ as shown in the same Section.

The first-step estimation equations are obtained as follows. Let us rewrite equation (42) at the uniform measurement times $i = 0, 1, ..., N-1$

$$\breve{x}_1(i) / \breve{x}_1(i) = -b_1\left(1 - \breve{x}_2(i) / \widehat{\overline{x}}_2\right)$$
$$\breve{x}_2(i) / x_2(i) = b_2\left(1 - \breve{x}_1(i) / \widehat{\overline{x}}_1\right) \quad , \quad (82)$$
$$\widehat{\overline{x}}_1 = \frac{1}{N}\sum_{i=0}^{N-1}\breve{x}_1(i), \; \widehat{\overline{x}}_2 = \frac{1}{N}\sum_{i=0}^{N-1}\breve{x}_2(i)$$

where the round mark $\cup$ denotes measurement and $\cap$ estimate, and the nonzero average $\widehat{\overline{x}}_k$ is estimated from the measurements. If the measurement noise $\tilde{x}_k(i)$ is zero mean, the estimate $\widehat{\overline{x}}_k, k = 1, 2$ is unbiased. Since the long-term averages of the velocity and of $\overline{x}_k - x_k(i)$ are zero, the measurement record must be split into two segments $\{\mathcal{S}_{k+}, \mathcal{S}_{k-}\}$ defined by

$$\mathcal{S}_{k+} = \left\{t_i; \breve{x}_k(i) > \overline{x}_k, i_0 \le i < i_1\right\}, \mathcal{S}_{k-} = \left\{t_i; \breve{x}_k(i) < \overline{x}_k, i_0 \le i < i_1\right\}, \quad (83)$$

where $i_0 = \gamma \widehat{P} / T > 0$ accounts for the estimation transient of $\breve{x}_k(i)$, $\widehat{P}$ is the estimated period and $\gamma = 0.5$ is assumed. The end time $i_1 = i_0 + m\widehat{P} / T$ selects an integer number of periods. The rate coefficients are estimated by separately computing the ratios between the partial sums over the positive and negative segments as follows

$$\widehat{b}_1 = -\left(\mu_{2+}\frac{\sum_{t_i \in \mathcal{S}_{2+}}\breve{x}_1(t_i) / \breve{x}_1(t_i)}{\sum_{t_i \in \mathcal{S}_{2+}}\left(1 - \breve{x}_2(t_i) / \widehat{\overline{x}}_2\right)} + \mu_{2-}\frac{\sum_{t_i \in \mathcal{S}_{2-}}\breve{x}_1(t_i) / \breve{x}_1(t_i)}{\sum_{t_i \in \mathcal{S}_{2-}}\left(1 - \breve{x}_2(t_i) / \widehat{\overline{x}}_2\right)}\right),$$
$$\widehat{b}_2 = \mu_{1+}\frac{\sum_{t_i \in \mathcal{S}_{1+}}\breve{x}_2(t_i) / \breve{x}_2(t_i)}{\sum_{t_i \in \mathcal{S}_{1+}}\left(1 - \breve{x}_1(t_i) / \widehat{\overline{x}}_1\right)} + \mu_{1-}\frac{\sum_{t_i \in \mathcal{S}_{1-}}\breve{x}_2(t_i) / \breve{x}_2(t_i)}{\sum_{t_i \in \mathcal{S}_{1-}}\left(1 - \breve{x}_1(t_i) / \widehat{\overline{x}}_1\right)} \quad (84)$$

where $\mu_{k+} + \mu_{k-} = 1$ accounts for the uncertainty of two terms in (84), and in absence of any other information $\mu_{k+} = \mu_{k-} = 1/2$. The above estimates can be proved to be unbiased, when measurement errors are zero mean.

The lower bound of the covariance $P(\widehat{\mathbf{p}}) = \mathcal{E}\left\{\left(\widehat{\mathbf{p}} - \mathcal{E}\{\widehat{\mathbf{p}}\}\right)\left(\widehat{\mathbf{p}} - \mathcal{E}\{\widehat{\mathbf{p}}\}\right)^T\right\}$ of the estimate $\widehat{\mathbf{p}}$ at the second step, may be predicted by the Cramèr-Rao bound $P(\mathbf{p})$ [17], where $\mathbf{p}$ is the true parameter and $P(\mathbf{p})$ holds

$$P(\mathbf{p}) = \left(\sum_{i=0}^{N-1} G(i, \mathbf{p})\tilde{S}^{-1}(i)G^T(i, \mathbf{p})\right)^{-1}. \quad (85)$$

In (85), $\tilde{S}(i) = \text{diag}\left(\tilde{\sigma}_1^2(i), \tilde{\sigma}_2^2(i)\right)$ is the covariance of the measurement error $\tilde{\mathbf{x}}(i) = [\tilde{x}_1, \tilde{x}_2](i)$ of the population measurements $\breve{\mathbf{x}}(i) = [\breve{x}_1, \breve{x}_2](i)$, under assumption of normality and statistical independence. The matrix $G(t, \mathbf{p})$ is the Jacobian matrix of the measurement model, i.e. of the free response of the LV equation (81). The columns of the Jacobian matrix

$$G(t,\mathbf{p}) = \left[ \frac{\partial x_1(t,\mathbf{p})}{\partial \mathbf{p}} \quad \frac{\partial x_2(t,\mathbf{p})}{\partial \mathbf{p}} \right] \tag{86}$$

are the solution of the linear time-varying state equations

$$\begin{aligned}
\frac{\partial \dot{x}_1(t)}{\partial \mathbf{p}} &= x_1 \left( \mathbf{g}_1(t) + a_{12} \frac{\partial x_2}{\partial \mathbf{p}} \right) + \left( -b_1 + a_{12} x_2 \right) \frac{\partial x_1}{\partial \mathbf{p}}, \quad \frac{\partial x_1(0)}{\partial \mathbf{p}} = \left[ 1, 0, 0, 0, 0, 0 \right] \\
\mathbf{g}_1(t) &= \left[ 0, -1, x_2, 0, 0, 0 \right] \\
\frac{\partial \dot{x}_2(t)}{\partial \mathbf{p}} &= x_2 \left( \mathbf{g}_2(t) - a_{21} \frac{\partial x_1}{\partial \mathbf{p}} \right) + \left( b_2 - a_{21} x_1 \right) \frac{\partial x_2}{\partial \mathbf{p}}, \quad \frac{\partial x_2(0)}{\partial \mathbf{p}} = \left[ 0, 0, 0, 1, 0, 0 \right] \\
\mathbf{g}_2(t) &= \left[ 0, 0, 0, 0, 1, -x_1(t) \right]
\end{aligned} \tag{87}$$

which are driven by $\mathbf{x}(t) = [x_1, x_2]$ and by the initial conditions.

### 3.3.2   Monte Carlo test

The first estimation step corresponding in (82) and (84) has been checked by a Monte Carlo trial of $m = 0, ..., M - 1$ runs [17]. The first, second and third estimation steps will be applied in Section 3.5. The discrete-time state equation of the third estimation step will be explained in Section 3.4. In the Monte Carlo test, the estimation of the initial conditions $\{\hat{x}_{10}, \hat{x}_{20}\}$ is assumed to be equal to the first measurement, i.e.

$$\{\hat{x}_{10}, \hat{x}_{20}\} = \{\bar{x}_1(0), \hat{x}_2(0)\}. \tag{88}$$

The measurement noise has been designed to be low frequency such to modulate the amplitude of the periodic LV response as as in Figure 10, left: the noise cutoff frequency $f_{wk} \cong 0.13$ Hz, $k = 1, 2$ is smaller than the response frequency $f_{LV} \cong 0.4$ Hz. Figure 10, left, shows the time profile of the deviation $\bar{\hat{x}}_k - \bar{x}_k(t_i)$, which is clearly disturbed by the low-frequency measurement noise. Figure 10, right, shows the time profile of volume rate error $\tilde{\dot{x}}(t_i)$. The volume rate reconstruction is essential for the first-step intrinsic rate estimation in (84).

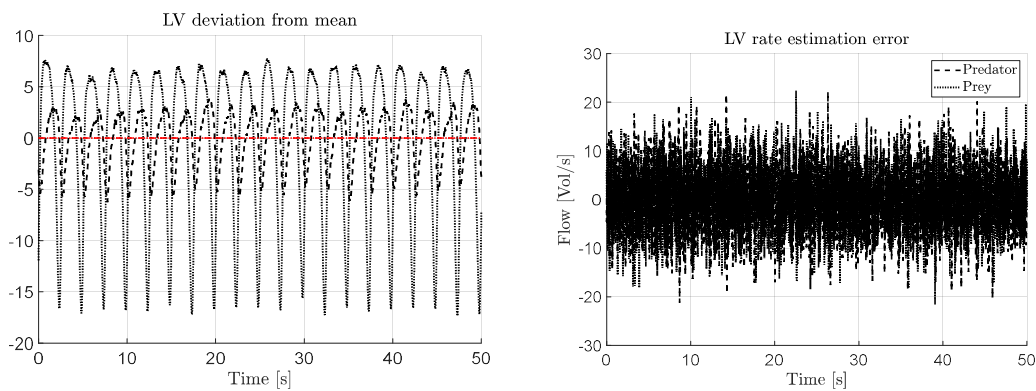

Figure 10. Left: deviation from the average population. Right: volume rate estimation error.

Table 2 shows test parameters and statistical results. They refer to normalized estimates, i.e. the estimated parameter $\hat{p}_k$ is divided by the true parameter value $p_k$, which implies unitary expected value:

$$\mathcal{E}\left\{\hat{p}_k / p_k\right\} = 1, \tag{89}$$

where $p_2 = b_1, p_3 = \underline{x}_1, p_5 = b_2, p_6 = \underline{x}_2$. Table 2 shows Monte Carlo bias $\delta_k$ and standard deviation $\sigma_k$ of each parameter, defined by

$$\delta_k = \frac{1}{M}\sum_{m=0}^{M-1}\left(\frac{\hat{p}_k(m)}{p_k} - 1\right) = \mu_k - 1$$

$$\sigma_k = \sqrt{\frac{1}{M}\sum_{m=0}^{M-1}\left(\frac{\hat{p}_k(m)}{p_k} - 1 - \delta_k\right)^2} \tag{90}$$

Table 2. Monte Carlo test parameters and statistics

| No | Parameter | Symbol | Unit | Value | Comments |
|----|-----------|--------|------|-------|----------|
| | Volume rate filter and measurement noise parameters | | | | |
| 1 | Estimated period | $\hat{P}$ | s | 2.5 (0.4 Hz) | |
| 2 | Measurement time unit | $T$ | s | 0.02 | |
| 3 | Volume rate filter eigenvalues | $\{\lambda_1, \lambda_2\}$ | | $\{0.5, 0.5\}$ | Discrete time |
| 4 | Number of used periods | | | 20 | |
| 5 | Low-frequency noise standard deviation | $\{\sigma_{w1}, \sigma_{w2}\}$ | Vol | $\{3,3\}$ | |
| 6 | Cutoff frequency | $\{f_{w1}, f_{w2}\}$ | Hz | $\{0.13, 0.13\}$ | |

| 7 | Measurement noise fractional bias | $\left\{\left|\mu_{w1}\right|/\overline{x}_1,\left|\mu_{w2}\right|/\overline{x}_2\right\}$ | fraction | $\{0.13,0.13\}$ | To avoid negative measurements |
|---|---|---|---|---|---|
| Monte Carlo test | | | | | |
| 8 | Number of runs | $M$ | | 100 | |
| 9 | Predator intrinsic rate $b_1$ | $\left\{\left|\delta_2\right|,\sigma_2\right\}$ | fraction | $\{0.054,0.045\}$ | |
| 10 | Predator average population $\underline{x}_1$ | $\left\{\left|\delta_3\right|,\sigma_3\right\}$ | fraction | $\{<0.001,0.015\}$ | |
| 11 | Prey intrinsic rate $b_2$ | $\left\{\left|\delta_5\right|,\sigma_5\right\}$ | fraction | $\{0.027,0.033\}$ | |
| 12 | Prey average population $\underline{x}_2$ | $\left\{\left|\delta_6\right|,\sigma_6\right\}$ | fraction | $\{<0.001,0.0066\}$ | |

Monte Carlo statistics in Table 2, rows 9 and 11, shows that the first-step estimation of the intrinsic rates is biased. This is due to the measurement error bias in Table 2, row 7.

Figure 11 shows the quantile-quantile (Q-Q) plot of the normalized estimate $\hat{b}_k(m)/b_k$, $k=1,2$, of the intrinsic rate coefficients versus the Gaussian variables $g_k(m/M)$ defined by

$$F\left(g_k;\mu_k,\sigma_k\right)=m/M,\qquad(91)$$

where $F\left(g_k;\mu_k,\sigma_k\right)=m/M$ is the cumulative Gaussian distribution defined by the estimation mean $\mu_k$ and standard deviation $\sigma_k$ in (90). The Monte Carlo estimates fit the Gaussian distribution.

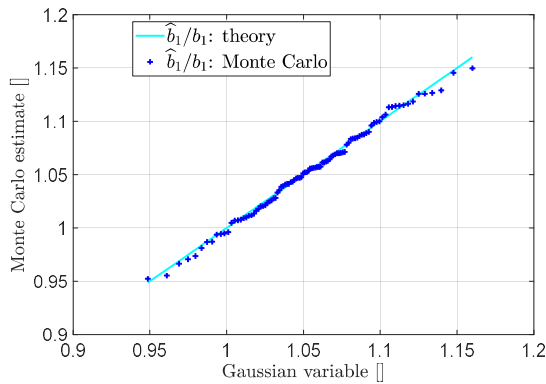
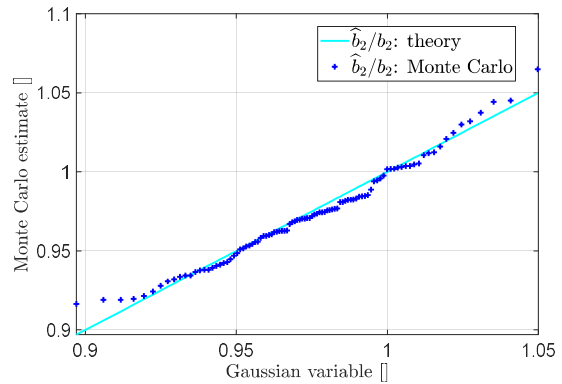

## 3.4 Discrete time LV equation

### 3.4.1 Forward Euler method

The simplest discretization method is the forward Euler method which amounts to replace the state derivative in (45) with

$$\dot{\mathbf{z}}(t) \cong \frac{\mathbf{z}(t+T) - \mathbf{z}(t)}{T}. \tag{92}$$

The corresponding DT equation from (45) holds

$$\mathbf{z}(i+1) = \mathbf{z}(i) + \begin{bmatrix} z_1(i) & 0 \\ 0 & z_2(i) \end{bmatrix} \begin{bmatrix} 0 & Tb_1 \\ -Tb_2 & 0 \end{bmatrix} (\mathbf{z}(i) - \overline{\mathbf{z}}), \ \mathbf{z}(0) = \mathbf{z}_0. \tag{93}$$

and has the same equilibrium points of the CT equation, namely $\overline{\mathbf{z}}_0 = [0,0]$ and the critical point $\overline{\mathbf{z}} = [1,1]$. The stability of the trajectories around the critical point depends on $T$ as follows. To this end, let us consider the perturbation $\delta\mathbf{z} = \mathbf{z} - [1,1]$ and the corresponding LTI equation:

$$\delta\mathbf{z}(i+1) = \delta\mathbf{z}(i) + T \begin{bmatrix} 1+\delta z_1(i) & 0 \\ 0 & 1+\delta z_2(i) \end{bmatrix} \begin{bmatrix} 0 & b_1 \\ -b_2 & 0 \end{bmatrix} \delta\mathbf{z}(i), \ \delta\mathbf{x}(0) = \delta\mathbf{x}_0. \tag{94}$$

The characteristic equation of the LTI part of the state matrix in (94) is the following

$$\lambda^2 - 2\lambda + 1 + T^2 b_1 b_2 = 0 \Rightarrow |\lambda_{1,2}| = \sqrt{1 + T^2 b_1 b_2} > 1, \tag{95}$$

and shows that the DT equation is always unstable for $T > 0$.

### 3.4.2 Stabilization and frequency tuning

When a DT state equation is employed as a state predictor for prediction, control or identification purposes, open-loop stability is not essential, since the state predictor is the result of a feedback stabilization of the DT state equation itself, the feedback being driven by the sampled measurements of the real dynamic system (alternatively, of an accurate simulator). As a matter of fact, weaker feedback corrections will be applied to more accurate DT equations.

In the case of nonlinear state equations, no general rule exists for building an exact DT equation, i.e. such that, given the time unit $T$ and the initial condition $\mathbf{z}_0$, the solution $\mathbf{z}(i)$ of (93) is the same as the sampled solution $\mathbf{z}(iT)$ of (45) for $i \geq 0$. To reach this objective, we proceed in two steps.

1) *Stabilization* of the perturbed LTI equation. We find the sampled-data equation of the LTI part of (59), which amounts to

$$\delta\mathbf{z}(i+1) = \delta\mathbf{z}(i) + \begin{bmatrix} \cos(\omega_{LV}T)-1 & \sqrt{\dfrac{b_1}{b_2}}\sin(\omega_{LV}T) \\ -\sqrt{\dfrac{b_2}{b_1}}\sin(\omega_{LV}T) & \cos(\omega_{LV}T)-1 \end{bmatrix}\delta\mathbf{z}(i) = \left(I + A(T)\right)\delta\mathbf{z}(i). \quad (96)$$

We replace the LTI matrix in (93) with the matrix $A(T)$ in (96) in order to guarantee the same equilibrium points of (93) and the perturbation stability of (96). We should remark that the free response of (96) around the critical point is only *quasi-periodic*, since in general the period $P_{LV} = 2\pi / \omega_{LV}$ imposed by the eigenvalues, is not a multiple of the time unit $T$. The DT nonlinear equation holds

$$\mathbf{z}_{DT}(i+1) = \mathbf{z}_{DT}(i) + \begin{bmatrix} z_1(i) & 0 \\ 0 & z_2(i) \end{bmatrix} A(T)\left(\mathbf{z}_{DT}(i)-\overline{\mathbf{z}}\right), \ \mathbf{z}_{DT}(0) = \mathbf{z}_{DT,0}. \quad (97)$$

$$x_{DT,k}(i) = z_{DT,k}(i)\overline{x}_k, \ k=1,2$$

2) *Frequency tuning*. In general, equation (97), though stable, drifts in a bounded way (see Figure 12) from the CT equation (45), like a pair of sine functions tuned on slightly different frequencies. In other words, the error equation between CT and DT is Lyapunov unstable but Lagrange stable [17]. The drift is eliminated by adding to (97) an input $\mathbf{u}(i)$ which is synthesized as a dynamic feedback driven by the model error

$$\tilde{\mathbf{x}}_m(i) = \overline{\mathbf{x}}(t_i) - \mathbf{x}_{DT}(i). \quad (98)$$

The simplest dynamic feedback is first-order as follows

$$\mathbf{x}_u(i+1) = \mathbf{x}_u(i) + L_1\tilde{\mathbf{x}}_m(i), \ \mathbf{x}_u(0) = \mathbf{x}_{u0}$$
$$\mathbf{u}(i) = \mathbf{x}_u(i) + L_0\tilde{\mathbf{x}}_m(i). \quad (99)$$

The gain matrices $L_1 = \text{diag}(l_{11}, l_{12})$ and $L_2 = \text{diag}(l_{21}, l_{22})$ are designed to stabilize the unstable error equation, which can be written as

$$\tilde{\mathbf{x}}(i+1) = \tilde{\mathbf{x}}(i) + \mathbf{h}\left(\mathbf{x}(t_i), \tilde{\mathbf{x}}(t_i), \mathbf{x}_{DT}(i)\right) + \overline{X}\mathbf{u}(i), \ \tilde{\mathbf{x}}(0) = \tilde{\mathbf{x}}_0,$$
$$\tilde{\mathbf{x}}(i) = \overline{X}\mathbf{z}_{DT}(i) - \mathbf{x}(t_i), \ \overline{X} = \text{diag}(\overline{x}_1, \overline{x}_2) \quad (100)$$

where $\mathbf{h}(\cdot)$ is the perturbation to be cancelled.

Figure 12 shows the open-loop bounded drift of the error equation (100) affecting the model error $\tilde{\mathbf{x}}_m(i)$ in (98).

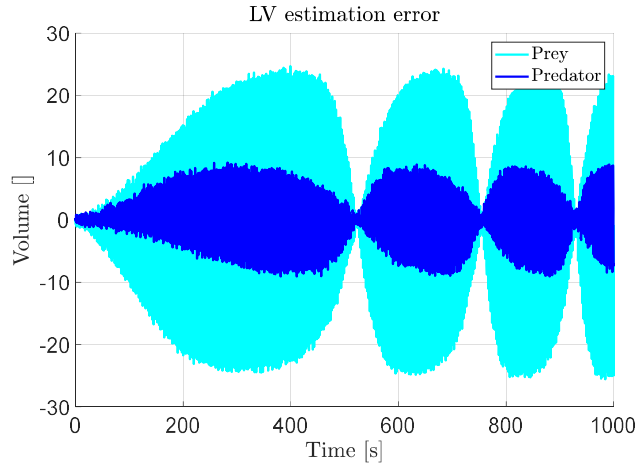

Figure 12. Bounded drifting error between CT equation and stable DT equation of the LV dynamics.

Figure 13 shows the LF components of $\mathbf{u}(i)$, i.e. the state vector $\mathbf{x}_u(i)$ in (99), in charge of cancelling $\mathbf{h}(\cdot)$ in (100) in the unnoisy (left) and noisy case (right). The closed loop frequency BW is 1 Hz.

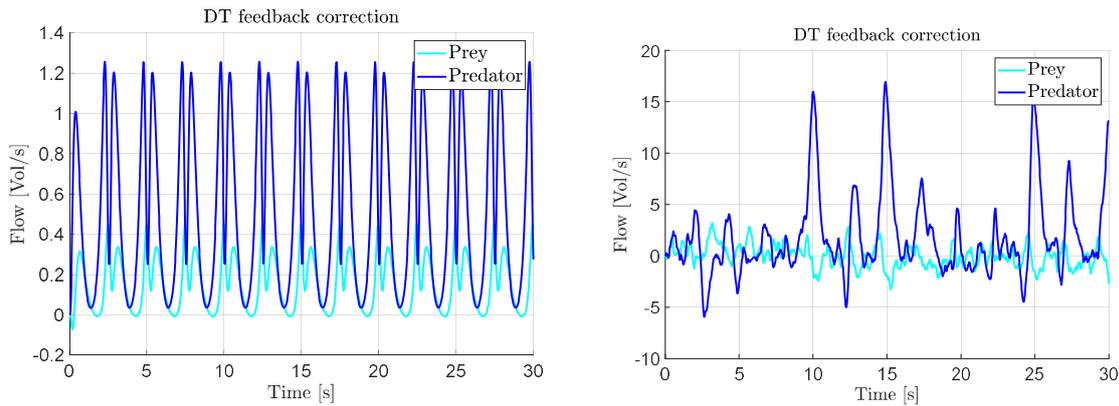

Figure 13. Left: periodic feedback without measurement noise. Right: random feedback in the noisy case.

## 3.5 Application of LV equation: snowshoe hare and Canadian lynx data

The parameter estimation method of Section 3.3 will be applied to a short-time period of the well-known 'snowshoe hare and Canadian lynx population data' collected in the northern Canadian forests from 1845 to 1933 [16]. The recorded population data were actually the number of furs caught by the Hudson Bay Company, and an underlying assumption of the data analysis is that the number of caught furs was proportional to the actual population of hares and lynxes. The raw data are limited to 20 years from 1900 to 1920 and denote the yearly average population. The raw data are firstly interpolated with a cubic spline to create a smooth profile, as it allows population rate computation. $N = 10$ interpolation points have been assigned to each year, which implies a time unit $T_s = 0.1\,\mathrm{y}$. The initial time $t_0$ has been set to zero, i.e. $t_0 = 0$. The data uncertainty due to measuring errors is not known. The time profiles of interpolated and raw data are reported in Figure 14. The population volume is expressed in kVol=1000 individuals.

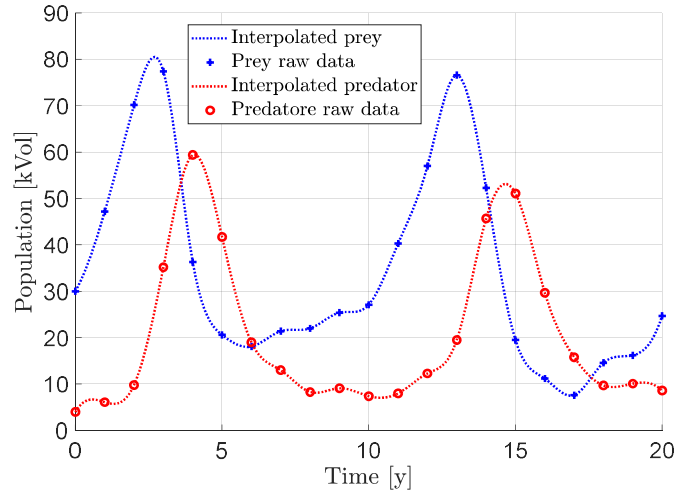

Figure 14. Spline interpolated and raw data (marked by crosses and circles).

The first and second step estimated parameters are reported in Table 3. The first-step estimated value is in bracket.

| No | Parameter | Symbol | Unit | Value | Comments |
|---|---|---|---|---|---|
| | Table 3. | | Hare and Lynx case parameter estimation | | |
| 1 | Lynx initial population | $x_1(0)$ | kVol | 4.04 (4.00) | |
| 2 | Hare initial population | $x_2(0)$ | kVol | 36.22 (30.00) | |
| 3 | Lynx rate coefficient | $b_1$ | kVol/y | 0.92 (0.73) | |
| 4 | Hare rate coefficient | $b_2$ | kVol/y | 0.48 (0.55) | |
| 5 | Lynx average population | $\overline{x}_1$ | kVol | 19.48 (20.86) | |
| 6 | Hare average population | $\overline{x}_2$ | kVol | 33.78 (34.35) | |
| 7 | LV period | $P_{LV}$ | y | 9.48 (9.88) | |

| 8 | Mean energy | $E$ | kVol/y | 0.38 (0.29) | |
| 9 | Residual RMS, lynx | $\tilde{\sigma}_{x1}$ | Vol/Vol | 0.14 (0.62) | Fractional error |
| 10 | Residual RMS, hare | $\tilde{\sigma}_{x2}$ | Vol/Vol | 0.13 (0.47) | same |
| 11 | Error correlation (angle) | $\rho_{12}$ | (rad) | -0.24 (1.81) | |

The second-step estimated time profile is compared to the interpolated raw data in Figure 15, left. The estimation residuals, normalized by the population averages, are shown in Figure 15, right. Hare and lynx residuals are close to be each other orthogonal, as proved by the small correlation in Table 3, row 11. The estimated profiles show a significant deviation from the interpolated raw data during the low population period of both hares and lynxes from year 5 to 13. We can only imagine an external cause, like for instance, a higher proportion of caught furs with respect to the actual population than otherwise, or a more complex competition dynamics as proved in [20] and [21].

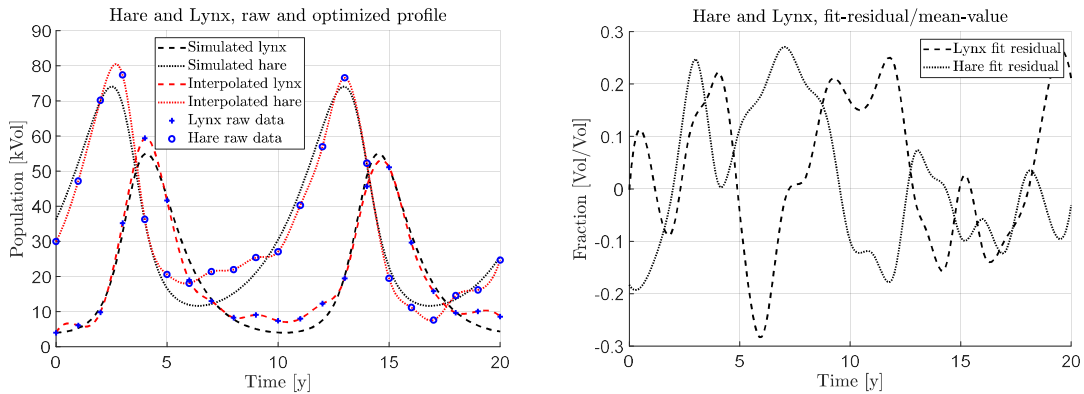

Figure 15.    Left: estimated population time profile compared to raw data. Right: fractional residuals.

The deviation the estimated profile from raw data is shown also by the plot, in Figure 16, of the energy $E$ defined in (50). The deviation from year 5 to year 13 is evident and justifies the higher energy of the estimated model (black line).

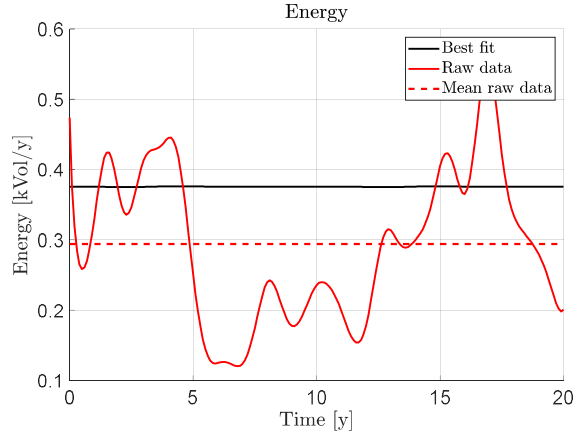

Figure 16.    Energy of the interpolated raw data (red) and the mean value (dashed red) compared with the energy of the estimated population profiles (black).

The third estimation step provides the input perturbation $u_k(t)$ in (81) as the combination of low (LF)-and high-frequency (HF) components. The corresponding DT equation is as follows:

$$
\begin{aligned}
x_{uk}(i+1) &= x_{uk}(i) + w_{1k}(i), \ x_{uk}(0) = x_{uk0} \\
u_k(i) &= x_{uk}(i) + w_{0k}(i) \\
&\quad\text{LF}\qquad\text{HF\quad components}
\end{aligned}
\tag{101}
$$

where the feedback signal $w_{jk}(i), j = 0,1$ is proportional to the model error $\tilde{x}_k(i) = \bar{x}_k(i) - \hat{x}_k(i)$, and $\hat{x}_k(i)$ is the state variable of the DT LV equation to be explained below. The resulting PI (proportional and integrative) feedback driven by $\tilde{x}_k(i)$ has the form

$$
\begin{aligned}
x_{uk}(i+1) &= x_{uk}(i) + L_{1k}\tilde{x}_k(i), \ x_{uk}(0) = x_{uk0} \\
u_k(i) &= x_{uk}(i) + L_{0k}\tilde{x}_k(i)
\end{aligned}
\tag{102}
$$

where $\{L_{0k}, L_{1k}\}$ are feedback gains fixing the eigenvalues (and the frequency BW) of the closed-loop system made by the DT LV equation and (102). As we will see below, the accuracy of the DT LV equation depends on the time unit $T$, which has been fixed smaller than the interpolation unit $T_s$, i.e. $T = 0.2 T_s = 0.02$ y. The closed-loop BW $f_c$ has been fixed close to the interpolation Nyquist frequency, i.e. $f_c \cong 5$ Hz. The estimated DT input signals $u_k(i), k = 1, 2$, in [KVol/y] units, are shown in Figure 17. The residual normalized error $\tilde{x}_k / \bar{x}_k$ becomes negligible with respect to residuals in Figure 15, right The residula RMS is less than 0.001 Vol/Vol, a value to be compared with Table 3, rows 9 and 10.

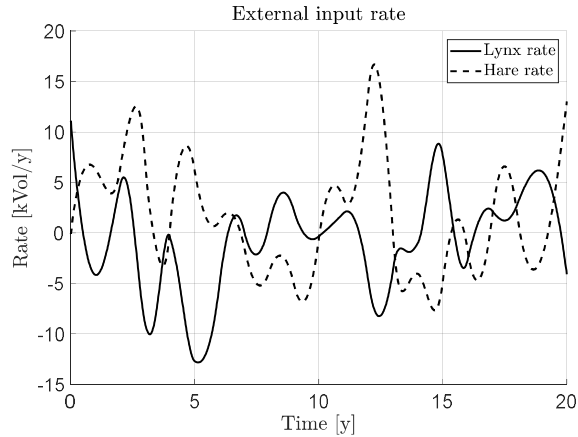

Figure 17.    Input signals (population rates) to LV equation to explain the residuals of the second step.

The time profiles in Figure 17 appear to be dominated by 2-year period fluctuations, which are visible also in Figure 15, right, and to a less extent in Figure 15, left. Likely, such fluctuations may be fit by a short-time, time-varying dynamics: cycle explanation and fitting is beyond the paper aim. The correlation between lynx and hare profiles is about -0.4, a bit larger in magnitude than the correlation in Table 3, rows 11.

The application of the two-species LV equation (7) to the lynx-hare competition has been judged too simple, from experimental data as in [20] and [21]. In essence, the hare is influenced by many predators other than lynx, and the lynx is primarily influenced by the snowshoe hare.

## 3.6    Application of LV equation: resource exploitation

### 3.6.1    A limiting form of the LV equation

In [2] a limiting form of the LV equations has been applied to resource exploitation and specifically to justify, under simple assumptions, and to generate the well-known bell-shaped Hubbert's curve, which has been obtained from the logistic equation in Section 2.1. The key difference from the logistic equation is the assumption of two competing species, the resource, playing the role of the 'prey', and the human capital, employed in exploitation, playing the role of 'predator'. Both resource and capital interact. The key assumption of the mathematical model is that the resource either cannot reproduce or the reproducing rate is negligible (non-renewable resource). Both capital and resource current amount (volume, stock) can be measured in currency units, energy units or mass units. Let us rewrite (42) with the external perturbation $u_k(t)$, already introduced in (81):

$$\text{capital: } \dot{x}_1(t) = -\left(b_1 - a_{12}x_2(t)\right)x_1(t) + u_1(t), \; x_1(0) = x_{10}, \; x_1(t) \geq 0$$
$$\text{resource: } \dot{x}_2(t) = \left(b_2 - a_{21}x_1(t)\right)x_2(t) + u_2(t), \; x_2(0) = x_{20}, \; x_2(t) \geq 0 \tag{103}$$

where the same unit [Vol] applies to the pair of state variables. The assumption that resource cannot reproduce by itself implies

$$b_2 \to 0 . \qquad (104)$$

By assuming zero perturbation $u_k(t) = 0$, the equilibrium points of (103) and (104) are as follows:

$$\overline{\mathbf{x}} = \left[ \overline{x}_1 = 0, \overline{x}_2 \geq 0 \right]. \qquad (105)$$

In other terms, any resource value is an equilibrium point. Among such points, the critical point is defined by $\overline{\mathbf{x}}_c = \left[ \overline{x}_{1c} = 0, \overline{x}_{2c} = b_1 / a_{12} \right]$. It is left to the reader to find the forced equilibrium points under $u_k(t) = \overline{u}_k \neq 0$. The local stability of the equilibrium points in (105) can be inferred from the perturbation equation of the state variable $\delta \mathbf{x} = \mathbf{x} - \overline{\mathbf{x}}$ and perturbation input $\mathbf{u} = [u_1, u_2]$, namely

$$\delta \dot{\mathbf{x}}(t) = \begin{bmatrix} -a_{11} = -b_1 \left(1 - \overline{x}_2 / \overline{x}_{2c}\right) & 0 \\ -a_{21}\overline{x}_2 & 0 \end{bmatrix} \delta \mathbf{x}(t) + \mathbf{u}(t), \delta \mathbf{x}(0) = \delta \mathbf{x}_0 . \qquad (106)$$

We remark that (106) is still nonlinear since perturbations from zero equilibrium cannot be negative. The zero equilibrium $\overline{\mathbf{x}}_0 = \left[ \overline{x}_1 = 0, \overline{x}_2 = 0 \right]$ is clearly stable since $\delta \mathbf{x}_0 > 0$. The stability of the nonzero equilibrium $\overline{\mathbf{x}} = \left[ \overline{x}_1 = 0, \overline{x}_2 > 0 \right]$ splits in two cases:

1) Instability, when $\overline{x}_2 / \overline{x}_{2c} \geq 1$. In the equality case, (106) becomes the series of two integrators.

2) Stability (not asymptotical stability), when $\overline{x}_2 / \overline{x}_{2c} < 1$. Moreover, since $\delta x_1(t) > 0$, the initial resource perturbation can only decrease as shown by the free response

$$\delta x_1(t) = \exp(-a_{11}t), \ \operatorname{sgn}(a_{11}) \ \text{any}$$
$$\delta x_2(t) = \delta x_{20} - a_{21}\overline{x}_2 \delta x_{10} \int_0^t \exp(-a_{11}\tau)\, d\tau \qquad (107)$$

The free response shape under $u_k(t) = 0$ of the nonlinear equation (103) can be inferred by solving separately the two equations as follows

$$\ln\left(x_1(t) / x_{10}\right) = -b_1 t \left(1 - \frac{1}{t}\int_0^t \frac{x_2(\tau)}{\overline{x}_2}\, d\tau\right)$$
$$\ln\left(x_2(t) / x_{20}\right) = -a_{21}\int_0^t x_1(\tau)\, d\tau x_{20} \qquad (108)$$

The resource volume $x_2(t)$ always decreases to the limiting value $x_{2\infty}$ in (110) below, whereas the capital volume increases for $x_2(t) > \overline{x}_2$ and then decreases to zero. As a result, the capital time profile $x_1(t)$ follows a bell-shaped profile. The non-negative resource production $p(t)$, defined by

$$p(t) = -\dot{x}(t) = a_{21}x_1(t)x_2(t), \tag{109}$$

follows the bell shape too, since $x_2(t)$ is a monotonic profile. Bell-shaped profiles can be considered as the limit of the LV periodic profiles, when the period $P_{LV} = 2\pi / \sqrt{b_1 b_2} \to \infty$, which occurs in this case, since $b_2 \to 0$ has been assumed. The resource volume $x_2(t)$ decreases to a limiting value $x_{2\infty}$, since the integral of the capital $x_1(t)$ in (108) is finite. From (108), $x_{2\infty}$ holds

$$x_{2\infty} = \exp\left(-a_{21}\int_0^\infty x_1(\tau)d\tau\right)x_{20}. \tag{110}$$

A further variable of interest as pointed out in [2] is the 'return on investment' (ROI) $q(t) = p(t)/x_1(t)$ $\left[\text{unit}^{-1}\right]$ which from (109) is proportional to the resource volume

$$q(t) = a_{21}x_2(t), \tag{111}$$

and, therefore, is always decreasing. As a further property, let us denote with $t_1$ the time when $x(t_1) = x_{10}$. From (108) we find the equivalent of (58) to the present aperiodic solution, namely:

$$\frac{1}{t_1}\int_0^{t_1} x_2(\sigma)d\sigma = \bar{x}_2 = b_1/a_{21}. \tag{112}$$

Typical time profiles are in Figure 18, left. The counter-clockwise phase diagram production-capital is in Figure 18, right. The equation parameters are

$$\left[x_{10}, x_{20}, b_1, a_{12}, a_{21}\right] = \left[0.5\times10^{-3}, 3.0, 1.0, 1.0, 0.5\right]. \tag{113}$$

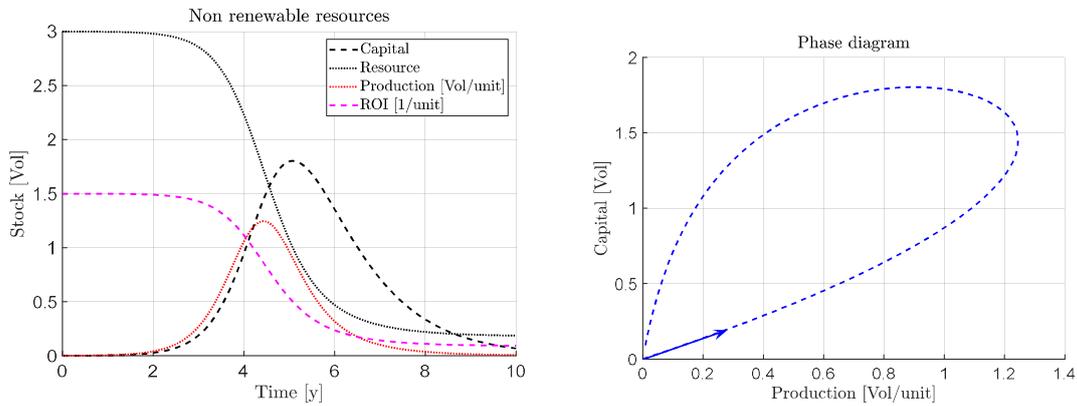

Figure 18.     Left: Time profiles of capital stock, resource stock, production and ROI. Right: production-capital phase diagram.

Since as shown by (118), the capital stock $x_1$ can be made function of the production $p$, a phase diagram as in Figure 18, right, can be plotted. The diagram, as indicated by the initial arrow, is counter-clockwise, which implies that the internal area has negative sign. The internal area is the result of the integral

$$V\left(0,t_1\right)=\int_{p(0)}^{p(t_2)} x_1\left(p\right)\dot{p}\left(\tau\right)d\tau=\int_{p(0)}^{p(t_2)} x_1\left(p\right)dp,\tag{114}$$

where $t_2$ is defined by $p\left(t_2\right)=p\left(0\right)$. To show that the integral in (114) is always negative let us define two other time instants: $t_c$ is defined by $x_2\left(t_c\right)=\underline{x}_2$ and corresponds to

$$t_c=\arg\max_{t\geq 0} x_1\left(t\right).\tag{115}$$

The second time instant $t_p$ is defined by

$$t_p=\arg\max_{t\geq 0} p\left(t\right).\tag{116}$$

The integral in (114) is always negative because $t_p<t_c$ as one can check from Figure 18, right. To prove it, let us set to zero the production time derivative which provides

$$\frac{\dot{p}\left(t\right)}{a_{21}}=\dot{x}_1 x_2-px_1=0 \Rightarrow p\left(t_p\right)=b_1\left(x_2\left(t_p\right)-\overline{x}_2\right)x_2\left(t_p\right)/\overline{x}_2,\tag{117}$$

and proves that $t_p<t_c$ since $p\left(t_p\right)>0$. In summary, model (103) with $b_2=0$, tells us that the capital stock is more efficiently employed when the resource stock is somewhat larger than the critical value $\overline{x}_c$, the limiting epoch being $t_p$. The same meaning is given by the ROI $q\left(t\right)$.

Though very simple and maybe questionable, equations (103) and (104) imply that the equilibrium capital $\overline{x}_1$ is zero as in (105). On the contrary, a nonzero equilibrium capital $\overline{x}_1=a_{21}/b_2$ would appear for $b_2>0$ as in the classical LV equation (42) or by assuming a nonzero mean perturbation $u_2\left(t\right)=\overline{u}_2>0$. This implies that renewable resources push a nonzero steady-state human capital!

### 3.6.2 A first-order equation of the capital stock driven by production

If production $p\left(t\right)=-\dot{x}_2\left(t\right)$ is known instead of the resource stock $x_2\left(t\right)$, the state equation (103) becomes first-order and LTI as follows

$$\begin{aligned}\dot{x}_1\left(t\right)&=-b_1 x_1\left(t\right)+a_1 p\left(t\right)+u_1\left(t\right),\ x_1\left(0\right)=x_{10},\ x_1\left(t\right)\geq 0\\ a_1&=a_{12}/a_{21}\end{aligned},\tag{118}$$

where $a_1$ expresses how resources are transformed into capital. The solution of (118) is explicit as follows

$$x_1\left(t\right)=\exp\left(-b_1 t\right)x_{10}+\int_0^t \exp\left(-b_1\left(t-\tau\right)\right)\left(a_1 p\left(\tau\right)+u_1\left(\tau\right)\right)d\tau.\tag{119}$$

The corresponding sampled-data equation in the difference form, to be used for the parameter estimation, holds

$$\Delta x_1(i+1) = x_1(i+1) - x_1(i) = -\alpha x_1(i) + \beta p(i), \; x_1(0) = x_{10}$$
$$\alpha = 1 - \exp(-b_1 T), \; \beta = \frac{a_1}{b_1}(1 - \exp(-b_1 T)) \qquad , \qquad (120)$$

where $u_1(i) = 0$ has been assumed. The available measurements are given by

$$\bar{x}_1(i) = x_1(i) + \tilde{x}_1(i), \; i = 0, ..., N-1$$
$$\bar{p}(i) = p(i) + \tilde{p}(i) \qquad , \qquad (121)$$

where $\{\tilde{x}_1(i), \tilde{p}(i)\}$ denote the measurement noise. Parameter estimation assumes (120) and (121), but exploits the orthogonality of the pair $\{x_1, \Delta x_1 / T\}$ to implement a well-conditioned Gramian matrix. To this end, the measurement equation is written as

$$\bar{p}(i) = \gamma_1 \bar{x}_1(i) + \gamma_2 \Delta \bar{x}_1(i+1)/T + \eta(i)$$
$$\gamma_1 = \frac{\alpha}{\beta} = \frac{b_1}{a_1}, \gamma_2 = \frac{T}{\beta} = \frac{b_1}{a_1} \frac{T}{1 - \exp(-b_1 T)} \qquad , \qquad (122)$$

where $\eta(i) = \tilde{p}(i) - (\gamma_1 \tilde{x}_1(i) + \gamma_2 \Delta \tilde{x}_1(i+1)/T)$ is the measurement error, and where the parameter conversion from $\{\gamma_1, \gamma_2\}$ to $\{b_1, a_1\}$ holds

$$b_1 = -\ln(1 - \gamma_1 T / \gamma_2)/T, \; a_1 = b_1 / \gamma_1. \qquad (123)$$

Also in presence of a zero-mean measurement error, the estimate $\hat{\gamma} = [\hat{\gamma}_1, \hat{\gamma}_2]$ is biased, because of the noisy regression variables $\{\tilde{x}_1(i), \Delta \tilde{x}(i+1)/T\}$. Since the bias depends on the covariance of the measurement error, which is a second-order term in the expansion of $\hat{\gamma}$, we can assume $\mathcal{E}\{\hat{\gamma}\} \cong \gamma$. Given the covariance $\bar{P}_\gamma = \mathcal{E}(\tilde{\gamma}\tilde{\gamma}^T) = [\tilde{\sigma}_1^2, \tilde{\rho}_{12}\tilde{\sigma}_1\tilde{\sigma}_2, \tilde{\sigma}_2^2]$ of the estimation error $\tilde{\gamma} = \hat{\gamma} - \gamma$, the error covariance of the pair $\{\hat{b}_1, \hat{a}_1\}$ can be computed from the following expansion in terms of the components of $\tilde{\gamma}$:

$$\hat{b}_1 = b_1 + \hat{b}_1 \Rightarrow \tilde{b}_1 = \frac{\tilde{\gamma}_1/\gamma_1 - \tilde{\gamma}_2/\gamma_2}{\gamma_2 - \gamma_1 T}\gamma_1$$
$$\hat{a}_1 = a_1 + \tilde{a}_1 \Rightarrow \tilde{a}_1 = \frac{\tilde{b}_1}{\gamma_1} - \frac{b_1}{\gamma_1}\frac{\tilde{\gamma}_1}{\gamma_1} = \frac{\tilde{\gamma}_1/\gamma_1 - \tilde{\gamma}_2/\gamma_2}{\gamma_2 - \gamma_1 T} - a_1 \frac{\tilde{\gamma}_1}{\gamma_1} \qquad (124)$$

Table 4 and Figure 19 show estimation results of the model in (113). The measured variables are the capital stock and the production. The measurement error is a coloured noise with a driving noise pair with standard deviation $\{\sigma_{x1}, \sigma_p\}$ and cutoff frequency $\{f_{x1}, f_p\}$ as reported in Table 4. The last row of Table 4 reports the relative error RMS of the production estimation, defined by

$$\partial \hat{\sigma}_p = \frac{1}{\sum_{i=0}^{N-1} \bar{p}(i)} \sqrt{\frac{1}{N}\sum_{i=0}^{N-1}(p(i) - \hat{p}(i))^2} \; . \qquad (125)$$

| No | Parameter | Symbol | Unit | Value | Comments |
|---|---|---|---|---|---|
| | Table 4. | Test case parameters | | | |
| 1 | Estimated interval | $H$ | unit | 10 | |
| 2 | Measurement time unit | $T$ | unit | 0.05 | |
| 3 | Low-frequency noise standard deviation | $\{\sigma_{x1}, \sigma_p\}$ | Vol | $\{0.25, 0.1\}$ | (capital stock, production) |
| 4 | Cutoff frequency | $\{f_{x1}, f_p\}$ | 1/unit | $\{0.03, 0.03\}$ | |
| 5 | Capital obsolescence rate | $|\partial b_1|$ | fraction | <0.02 | |
| 6 | Resource to capital transformation | $|\partial a_1|$ | fraction | <0.03 | |
| 7 | Production estimation relative error (RMS) | $\partial\tilde{\sigma}_p$ | fraction | 0.09 | |

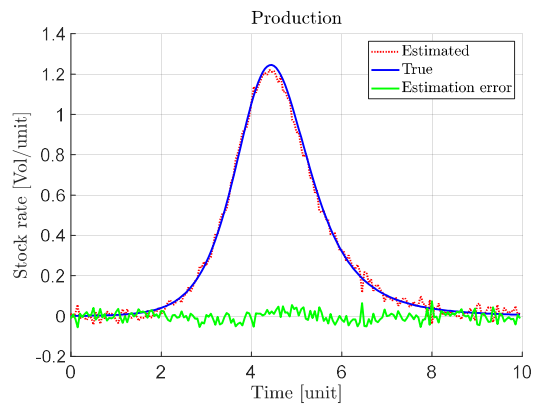

Figure 19.    True production, estimation and estimation error.

### 3.6.3   Application to experimental data: California gold rush

In [2], several fitting examples from experimental data have been proposed. We only consider the scarce data of the California gold rush accounted for in [18] and [19].

The gold rush (1848-1955) began on January 24, 1848, when flakes of gold were found in the American River by J.W. Marshall, a carpenter working to build a water-powered saw mill, the Sutter's Mill, near Coloma, California. In the March 1848, the discovery was announced through the streets of San Francisco by the newspaper publisher and merchant S. Brennan, who holding a vial of gold, walked along the streets shouting 'Gold, gold, gold from the American River'. The US-wide spread news attracted from 1848 to 1855 about 300000 people from the rest of United States and abroad.

Production and capital stock raw data are available from [18] and [19]. They are rather scarce, sparse and uncertain. Gold production $\bar{p}(i)$ is measured in MUSdollar/y=$10^6$USdollar/y. Capital stock $\bar{x}_1(i)$ is assumed to be proportional to the estimated number of gold prospectors [kVol]. Production and capital raw data have been spline interpolated with a time unit of $T = 0.1\,\text{y}$, beginning in 1843 and assuming zero production from 1843 to 1847. Raw and interpolated data are assumed to satisfy the DT equation (120) which has been converted into (122) for estimation purposes.

No measurement error is known. The least-squares solution of (122) is reported in Table 5. The capital obsolescence was found to be $0.75\,\text{y}^{-1}$, which means a reduction of about 50% per year. The resource to capital transformation was estimated to be about 1.6 prospector per 1000 US dollars, which means that in average each prospector produced about 625 US dollars per year. By assuming a price of about 0.75 US dollars each gram of gold, the average prospector produced less than 1 kg of gold per year.

| No | Parameter | Symbol | Unit | Value | Comments |
|----|-----------|--------|------|-------|----------|
| | Table 5. California gold rush parameters | | | | |
| 1 | Estimation interval | $H$ | y | 30 | 1943-1973 |
| 2 | Interpolation time unit | $T$ | y | 0.1 | |
| 3 | Capital obsolescence rate | $\hat{b}_1$ | 1/y | 0.75 | |
| 4 | Resource to capital transformation | $\hat{a}_1$ | Vol/USDollar | 0.016 | 1 Vol=1 prospector |
| 5 | Production estimation error (RMS) | $\tilde{\sigma}_p$ | $10^6$USDollar/y | 4.9 | |
| 6 | Relative estimation error | $\tilde{\sigma}_p / \bar{p}$ | fraction | 0.16 | |

Notwithstanding the sparse and uncertain data, the interpolated capital profile (the blue curve in Figure 20, left) seems capable of fitting the interpolated gold production except during the peak production around the 1852 year and during the stabilized production set up after the gold rush, from about 1865. The peak production around the 1852 year may be the result of improvements of the gold-recovery techniques from placer mining to hydraulic mining. In fact, by assuming the smooth capital curve in Figure 20, left, a production peak higher than the estimated profile should imply that other capitals than prospectors were employed.

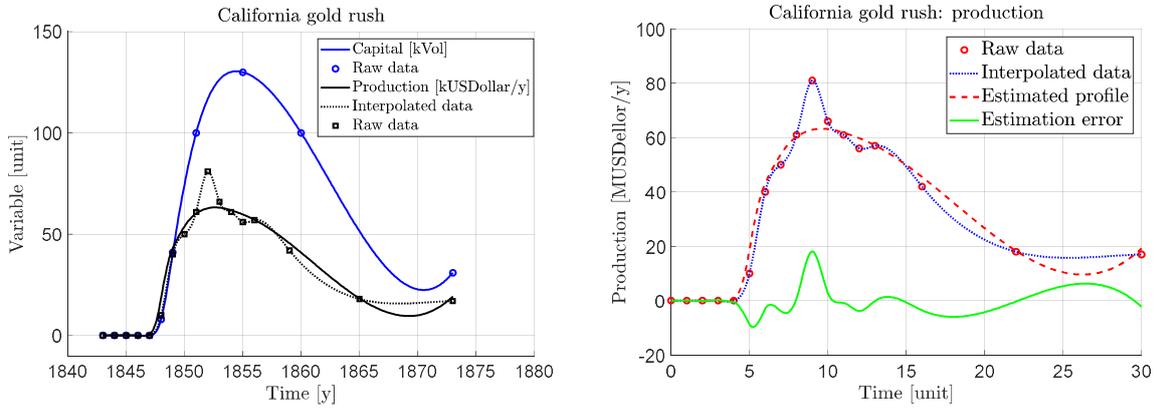

Figure 20.    Left: Capital and production raw data and estimated production. Right: production raw and interpolated data, estimated profile and estimation error.

By assuming the measurement error variance to be of the order of $\tilde{\sigma}_p^2$ in Table 5, row 5, the fractional standard deviation of $\left\{\tilde{b}_1, \tilde{a}_1\right\}$ was estimated from (124) to be less than 10%. As expected, the capital time profile lags the production profile of about two years, if measured between the peak values of the estimated profiles.

# 4 Riferences

# 5  Appendix

## 5.1  Poincaré maps

Given an autonomous state equation of order $n$

$$\dot{\mathbf{x}}(t) = \mathbf{f}(\mathbf{x}(t)), \ \mathbf{x}(0) = \mathbf{x}_0, \ \dim \mathbf{x} = n, \quad (126)$$

a way to find the properties of the trajectories in the phase space $\mathscr{X} \subseteq \mathbb{R}^n$ is to define a surface $\varSigma \subset \mathscr{X}$ of dimension $n-1$. Given an initial state $\mathbf{x}_0 \in \varSigma$ corresponding to a trajectory intersection, defined by $\left| \mathbf{f}(\mathbf{x}_0) \cdot \mathbf{n}((\mathbf{x}_0)) \right| > 0$ where $\mathbf{n}((\mathbf{x}_0))$ is the normal of the surface tangent Figure 21 plane in $\mathbf{x}_0$, find the trajectory intersections $\{\mathbf{x}_1, \mathbf{x}_2, \dots \in \varSigma\}$ with the surface itself at times $\{t_1 < t_2 < \dots\}$, such that

$$\text{sgn}\left(\mathbf{f}(\mathbf{x}_k) \cdot \mathbf{n}((\mathbf{x}_k))\right) = \text{sgn}\left(\mathbf{f}(\mathbf{x}_0) \cdot \mathbf{n}((\mathbf{x}_0))\right), k > 0, \quad (127)$$

i.e. surface crossing occurs in the same direction (see Figure 21). The *Poincaré or return map* $P : \mathscr{X} \rightarrow \mathscr{X}$ is the map that relates a generic initial state to the next return state

$$\mathbf{x}_{k+1} = P(\mathbf{x}_k), \ t_{k+1} > t_k. \quad (128)$$

The map does not exist if the trajectory does not return to the surface. There may exists trajectories which does not cross or just touch the surface. By excluding pathological trajectories, if the *trajectory is close (orbit)* we have

$$\mathbf{x}_{k+1} = \mathbf{x}_k \Rightarrow P(\mathbf{x}_k) = \mathbf{x}_k \quad (129)$$

and the time interval $T_k = t_{k+1} - t_k$ is the period of the orbit. In this case we may just restrict to the initial state $\mathbf{x}_0$ and to the first return state $\mathbf{x}_1 = \mathbf{x}_0$. A typical return surface is a $n-1$ dimensional hyperplane defined by

$$\varSigma = \left\{ \mathbf{x} \in \mathscr{X}. x_j(t) = x_{j0} = \text{constant} \right\}. \quad (130)$$

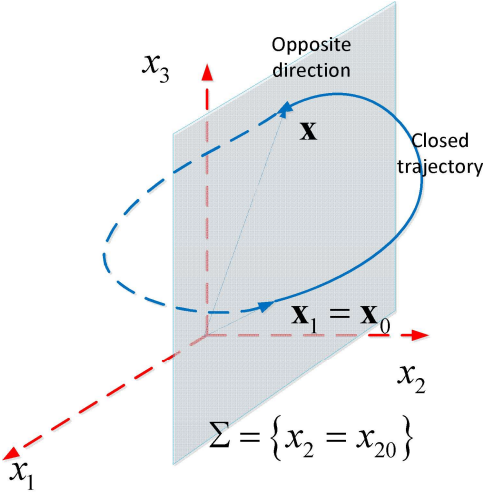

Figure 21.    A closed trajectory and the return surface.